\documentclass[
reprint,           
superscriptaddress,
amsmath,           
amssymb,           
aps,               
prl,               
notitlepage,       
floatfix,          
nofootinbib,
]{revtex4-1}

\usepackage{tensor}     
\usepackage{graphicx}   
\usepackage[
colorlinks=true,        
citecolor=blue,         
linkcolor=blue,         
urlcolor=blue           
]{hyperref}             
\usepackage{bm}         
\usepackage{xcolor}     
\usepackage{tabulary}
\usepackage{color}      
\usepackage[utf8]{inputenc} 
\usepackage[section]{placeins} 

\newcommand{\nc}{\newcommand*} 

\nc{\al}{\alpha}
\nc{\s}{\sigma}
\nc{\dt}{\delta}
\nc{\Dt}{\Delta}
\nc{\Ld}{\Lambda}
\nc{\p}{\partial}
\nc{\om}{\omega}
\nc{\Om}{\Omega}
\nc{\rd}{\mathrm{d}}
\nc{\Od}[1]{\mathcal{O}(#1)} 
\nc{\kp}{\kappa}
\nc{\one}{\uppercase\expandafter{\romannumeral1}}
\nc{\two}{\uppercase\expandafter{\romannumeral2}}
\nc{\three}{\uppercase\expandafter{\romannumeral3}}
\def\({\left(}
\def\){\right)}
\def\[{\left[}
\def\]{\right]}
\def\e{\begin{equation}}
\def\q{\end{equation}}
\def\m{\begin{eqnarray}}
\def\n{\end{eqnarray}}
\nc{\Eq}[1]{Eq.~\eqref{#1}}     
\nc{\Fig}[1]{Fig.~\ref{#1}}     
\nc{\Table}[1]{Table~\ref{#1}}  
\nc{\Sec}[1]{Sec.~\ref{#1}}     
\nc{\Msun}{M_\odot}             
\nc{\fpbh}{f_{\mathrm{pbh}}}    
\nc{\fpbhn}{f_{\mathrm{pbh0}}}    
\nc{\mR}{\mathcal{R}} 
\nc{\seq}{\sigma_{\mathrm{eq}}}
\nc{\ogw}{\Omega_{\mathrm{GW}}}
\nc{\gpcyr}{\mathrm{Gpc}^{-3}\,\mathrm{yr}^{-1}}
\nc{\lvc}{LIGO/Virgo} 
\nc{\SNR}{\mathrm{SNR}} 
\nc{\mmin}{{m_{\mathrm{min}}}}
\nc{\mmax}{{m_{\mathrm{max}}}}
\nc{\Mmin}{{M_{\mathrm{min}}}}
\nc{\fmin}{{f_{\mathrm{min}}}}
\nc{\VT}{\mathrm{VT}}
\nc{\rhoGW}{\rho_{\mathrm{GW}}}
\nc{\vth}{\vec{\theta}}
\nc{\vd}{\vec{d}}
\nc{\vla}{\vec{\lambda}}
\nc{\Nobs}{N_{\mathrm{obs}}}
\nc{\av}[1]{\langle #1 \rangle} 
\nc{\km}{\mathrm{km}}
\nc{\Mpc}{\mathrm{Mpc}}
\nc{\Tobs}{T_{\mathrm{obs}}}
\nc{\Ntemp}{N_{\mathrm{temp}}}
\nc{\addref}{[\textcolor{red}{add ref}] } 
\nc{\eg}{\textit{e.g.~}}
\nc{\app}{\approx}
\nc{\hf}{\frac{1}{2}}
\nc{\discuss}{\textcolor{red}{Add discussion here!}}
\nc{\red}[1]{\textcolor{red}{#1}}
\nc{\mH}{\mathcal{H}}
\nc{\cs}{c_s^2}
\nc{\Sij}[1]{S_{ij}^{(#1)}}
\nc{\vi}[1]{v_i^{(#1)}}
\nc{\no}{\nonumber}
\def\<{\left\langle}
\def\>{\right\rangle}

\nc{\bk}{\bm{k}}
\nc{\bq}{\bm{q}}
\nc{\bp}{\bm{p}}
\nc{\bl}{\bm{l}}
\nc{\bx}{\bm{x}}
\nc{\be}{\mathbf{e}}
\nc{\mS}{\mathcal{S}}
\nc{\te}{\tilde{\eta}}
\nc{\tp}{\tilde{p}}
\nc{\tk}{\tilde{k}}
\nc{\tx}{\tilde{x}}
\nc{\tF}{\tilde{F}}
\nc{\tA}{\tilde{A}}
\nc{\mkpq}{|\bk-\bp-\bq|}
\nc{\mpq}{|\bp-\bq|}
\nc{\mkp}{|\bk-\bp|}
\nc{\mSi}[1]{\mS^{(#1)}({\bk, \eta})}
\nc{\vk}{\vec{k}}
\nc{\kstar}{k_*}
\nc{\xstar}{x_*}
\nc{\mpbh}{m_{\rm{pbh}}}

\renewcommand{\vec}[1]{\boldsymbol{#1}} 

\begin{document}
	
\title{Log-dependent slope of scalar induced gravitational waves in the infrared regions}
	
\author{Chen Yuan}
\email{yuanchen@itp.ac.cn}
\affiliation{CAS Key Laboratory of Theoretical Physics, 
Institute of Theoretical Physics, Chinese Academy of Sciences,
Beijing 100190, China}
\affiliation{School of Physical Sciences, 
University of Chinese Academy of Sciences, 
No. 19A Yuquan Road, Beijing 100049, China}
	
\author{Zu-Cheng Chen}
\email{chenzucheng@itp.ac.cn} 
\affiliation{CAS Key Laboratory of Theoretical Physics, 
Institute of Theoretical Physics, Chinese Academy of Sciences,
Beijing 100190, China}
\affiliation{School of Physical Sciences, 
University of Chinese Academy of Sciences, 
No. 19A Yuquan Road, Beijing 100049, China}

\author{Qing-Guo Huang}
\email{huangqg@itp.ac.cn}
\affiliation{CAS Key Laboratory of Theoretical Physics, 
Institute of Theoretical Physics, Chinese Academy of Sciences,
Beijing 100190, China}
\affiliation{School of Physical Sciences, 
University of Chinese Academy of Sciences, 
No. 19A Yuquan Road, Beijing 100049, China}
\affiliation{Center for Gravitation and Cosmology, 
College of Physical Science and Technology, 
Yangzhou University, Yangzhou 225009, China}
\affiliation{Synergetic Innovation Center for Quantum Effects and Applications, 
Hunan Normal University, Changsha 410081, China}
	
\date{\today}

\begin{abstract}

We analytically calculate the scalar induced gravitational waves (SIGWs) and find a log-dependent slope of SIGW in the infrared regions $(f<f_c)$, namely $n_{\mathrm{GW}}(f)=3-2/\ln(f_c/f)$, and $n_{\mathrm{GW}}(f)=2-2/\ln(f_c/f)$ near the peak if the power spectrum of scalar curvature perturbation is quite narrow, where $f_c$ is roughly the peak frequency. Such a log-dependent slope can be taken as a new template for distinguishing SIGW from other sources. 


\end{abstract}
	
\pacs{???}
	
\maketitle
	
Dark matter (DM) is one of the components in the Universe and it makes up around $26\%$ of the total energy density at present \cite{Aghanim:2018eyx}. However, the nature of DM remains completely unknown. Even though there is a miracle for the weakly-interacting massive particles (WIMPs), the limits on them are tightening. Considering some alternative models to WIMPs becomes more and more important. Among the alternative models in literature, the primordial black holes (PBHs) have attracted much attentions in the past few years, in particular after the discovery of the gravitational waves (GWs) from the coalescence of a binary black hole by aLIGO \cite{Abbott:2016blz} because PBHs are supposed to provide a possible explanation if the abundance of stellar-mass PBHs in DM is roughly ${\cal O}(10^{-3})$ \cite{Sasaki:2016jop,Chen:2018czv,Chen:2019irf,Chen:2018rzo}. 
Up to now, there are a various observations which have put constraints on the abundance of PBH DM \cite{Tisserand:2006zx,Carr:2009jm,Barnacka:2012bm,Griest:2013esa,Graham:2015apa,Brandt:2016aco,Chen:2016pud,Wang:2016ana,Gaggero:2016dpq,Ali-Haimoud:2016mbv,Blum:2016cjs,Horowitz:2016lib,Niikura:2017zjd,Zumalacarregui:2017qqd,Abbott:2018oah,Magee:2018opb,Chen:2018rzo,Niikura:2019kqi,Chen:2019irf,Authors:2019qbw,Wang:2019kzb,Wu:2020drm}, but a substantial open window in the mass range of $[10^{-16},10^{-14}] \cup [10^{-13},10^{-12}] M_\odot$ is still allowed for PBHs composing of all of DM. See a recent summary in \cite{Chen:2019irf}.

PBHs are supposed to form from the gravitational collapse of over-densed regions seeded by relatively large curvature perturbations \cite{Hawking:1971ei,Carr:1974nx} on small scales which are less constrained by the cosmic microwave background (CMB) and large-scale structure observations. These over-densed regions are produced when curvature perturbations exceed a critical value and will collapse to form a PBH at about horizon size after the corresponding wavelength re-enters the horizon. 
However, how to test the postulation of PBH DM is still an open question. 
Actually, the curvature perturbations couple to the tensor perturbations at second-order, thus inevitably generating the scalar induced GWs (SIGWs) in the radiation dominated era \cite{tomita1967non,Matarrese:1992rp,Matarrese:1993zf,Matarrese:1997ay,Noh:2004bc,Carbone:2004iv,Nakamura:2004rm}. The enhancement of scalar curvature perturbations for significantly forming PBHs will generate relatively large SIGWs which provide a new way to probe PBHs \cite{Yuan:2019udt}. See some other related works in \cite{Ananda:2006af,Baumann:2007zm,Saito:2008jc,Assadullahi:2009jc,Bugaev:2009zh,Saito:2009jt,Bugaev:2010bb,Nakama:2016enz,Nakama:2016gzw,Garcia-Bellido:2017aan,Sasaki:2018dmp,Espinosa:2018eve,Kohri:2018awv,Cai:2018dig,Bartolo:2018evs,Bartolo:2018rku,Unal:2018yaa,Inomata:2018epa,Clesse:2018ogk,Cai:2019amo,Inomata:2019zqy,Inomata:2019ivs,Cai:2019elf,Cai:2019cdl,Chen:2019xse,Yuan:2019fwv}. 

A normalized stochastic gravitational-wave background (SGWB) spectral energy density $\Omega_{\mathrm{GW}}(f)$ expresses the GW spectral energy density in terms of the energy density per logarithmic frequency interval divided by the cosmic closure density, namely \cite{Maggiore:1999vm,Thrane:2013oya}
\m
\Omega_{\mathrm{GW}}(f)\equiv {1\over \rho_c}{d\rho_{\mathrm{GW}}\over d\log f}={2\pi^2\over 3H_0^2}f^3 S_h(f), 
\n
where $\rho_{\mathrm{GW}}$ and $\rho_c$ are the energy density of GWs and critical density, $f$ is the GW frequency, $H_0$ is the Hubble constant, and $S_h(f)$ is the spectral density. Conventionally, $\Omega_{\mathrm{GW}}(f)$ is modeled as a power law form, i.e. 
\m
\Omega_{\mathrm{GW}}(f)\propto f^{n_{\mathrm{GW}}}
\n 
with a slope ${n_{\mathrm{GW}}}$. The predicted SGWB from compact binary coalescences is well modeled by a power law of slope ${n_{\mathrm{GW}}}=2/3$, and ${n_{\mathrm{GW}}}=0$ corresponds to a scale-invariant energy. For the primordial GWs, $\Omega_{\mathrm{GW}}(f) \propto f^{n_t+\alpha_t \ln (f/f_{\mathrm{CMB}})/2}$, where $n_t$ is the spectral index of primordial GW power spectrum and $\alpha_t$ is the running of spectral index \cite{Lasky:2015lej,Li:2019vlb,Li:2019efi,Li:2018iwg}. For the white noise which corresponds to a random signal, the spectral density $S_h$ is a constant, and then $\Omega_{\mathrm{GW}}(f)\propto f^3$.   
In this sense, the $f^3$ behavior seems trivial rather than a model-independent evidence for the detection of SIGWs. That is why there are many processes generating a SGWB scaling as $f^3$ in the infrared limit, such as GWs from first-order phase transition \cite{Caprini:2009fx} and GWs induced by an inflaton field \cite{Liddle:1999hq}. In addition, $\Omega_{\mathrm{GW}}(f)$ drops down quickly in the infrared region from the peak, making it difficult to detect the $\propto f^3$ slope. 

Even though the behavior of SIGW $\Omega_{\mathrm{GW}}(f)$ is expected to be dependent on the power spectrum of the scalar curvature perturbations, we find $n_{\mathrm{GW}}=3-2/\ln(f_c/f)$ for SIGW in the infrared region, and $n_{\mathrm{GW}}=2-2/\ln(f_c/f)$ near the peak if the scalar power spectrum is very narrow. In the infrared limit $(f\rightarrow 0)$, $n_{\mathrm{GW}}\rightarrow 3$, indicating that the correlation of perturbations can be neglected and the signal behaves randomly. This log-dependent slope can be taken as a distinguishing feature for the SIGWs.


The perturbed Friedmann-Robertson-Walker (FRW) metric for a perturbed universe in Newtonian gauge takes the form, \cite{Ananda:2006af}, 
\e
\rd s^{2}=a^{2}\left\{-(1+2 \phi) \rd \eta^{2}+\left[(1-2 \phi) \delta_{i j}+\frac{h_{i j}}{2}\right] \rd x^{i} \rd x^{j}\right\},
\q
where $\phi$ is the scalar perturbation and $h_{ij}$ is the GW perturbation. In a radiation dominated universe without entropy perturbations, the equation of motion for $\phi$ governed by Einstein equation reads  
\e
\phi_{\vec{k}}''(\eta)+{4\over\eta}\phi_{\vec{k}}'(\eta)+{k^2\over3}\phi_{\vec{k}}(\eta)=0
\q
in Fourier space and $k=2\pi f$. 
This equation of motion has an attenuation solution given by, \cite{Baumann:2007zm}, 
\e\label{phi}
\phi_{\vec{k}}(\eta) \equiv \phi_{\vec{k}}\frac{9}{(k\eta)^2} \[  \frac{\sin(k\eta/\sqrt{3})}{k\eta/\sqrt{3}} - \cos(k\eta/\sqrt{3})\],
\q
where $\phi_{\vec{k}}$ is the primordial perturbation whose value at $\eta=0$ is given by inflation models. The equation of motion for the tensor components is given by Einstein equation at second-order, namely 
\e\label{eqh}
h_{i j}^{\prime \prime}+2 \mathcal{H} h_{i j}^{\prime}-\nabla^{2} h_{i j}=-4 \mathcal{T}_{i j}^{\ell m} S_{\ell m},
\q
where the prime denotes the derivative with respect to the conformal time $\eta$, and $\mathcal{H}=a'/a$ is the conformal Hubble parameter. The source term \cite{Ananda:2006af}
\e
S_{ij}= 4 \phi \p_i\p_j\phi + 2\p_i\phi\p_j\phi-{1\over\mH^2}\p_i \(\mH\phi + {\phi'}\)\p_j\(\mH\phi + {\phi'}\),
\q
is projected to transverse-traceless gauge by the projection operator $\mathcal{T}_{i j}^{\ell m}$, i.e. 
\e
\mathcal{T}_{i j}^{\ell m}=\int {\mathrm{d}^3\vec{k}\over (2\pi)^{3/2}}\mathrm{e}^{i\vec{k}\cdot\vec{x}}\Big[\mathrm{e}_{ij}(\vec{k})\mathrm{e}^{lm}(\vec{k})+\bar{\mathrm{e}}_{ij}(\vec{k})\bar{\mathrm{e}}^{lm}(\vec{k})\Big].
\q
Here the two polarization tensors are defined by 
\e
\begin{split}
\mathrm{e}_{ij}(\vec{k})&\equiv{1\over\sqrt{2}}\[\mathrm{e}_i(\vec{k})\mathrm{e}_j(\vec{k})-\bar{\mathrm{e}}_i(\vec{k})\bar{\mathrm{e}}_{j}(\vec{k})\],\\
\bar{\mathrm{e}}_{ij}(\vec{k})&\equiv{1\over\sqrt{2}}\[\mathrm{e}_i(\vec{k})\bar{\mathrm{e}}_j(\vec{k})+\bar{\mathrm{e}}_i(\vec{k}){\mathrm{e}}_{j}(\vec{k})\],
\end{split}
\q
where $\mathrm{e}(\vec{k})$ and $\bar{\mathrm{e}}(\vec{k})$ are two time-independent unit vectors orthogonal to $\vec{k}$.
After solving Eq.~(\ref{eqh}) in Fourier space by Green's function method, one obtains, \cite{Baumann:2007zm}, 
\e\label{hsol} 
h(\vec{k},\eta) = \frac{1}{k a(\eta)} \int \rd \te \sin(k\eta - k\te) a (\te) \mathcal{S}_{\vec{k}}(\te),
\q
where $\mathcal{S}_{\vec{k}}(\eta)\equiv -4e^{ij}(\vec{k}) \tilde{S}_{ij}(\vec{k},\eta)$ with $\tilde{S}_{ij}(\vec{k},\eta)$ to be the Fourier transformed source term. Then
the dimensionless power spectrum of the SIGWs, $\mathcal{P}_h(k)$, can be evaluated by the two point correlation 
\e\label{Ph}
\left\langle h(\vec{k},\eta) h(\vec{k'},\eta)\right\rangle\equiv\frac{2 \pi^{2}}{k^{3}} \mathcal{P}_{h}(k,\eta) \delta(\vec{k}+\vec{k'}).
\q
And then, 	at the matter-radiation equality, 
\e\label{omiga}
\Omega_{\mathrm{GW,eq}}(k)= \frac{1}{24} \(\frac{k}{\mH}\)^2 \overline{\mathcal{P}_{h}(k, \eta)},
\q
where we have summed over two polarization modes and take the oscillation average. The present density parameter can be evaluated by, \cite{Espinosa:2018eve,Kohri:2018awv}, 
\m\label{omegagw}
\Omega_{\mathrm{GW}}(k)&&=\Omega_{r}\times\Omega_{\mathrm{GW,eq}}(k)=\Omega_{r}\times\Omega_{\mathrm{GW}}(\eta\rightarrow\infty,k)\no\\
&&={\Omega_{r}}\int_{0}^{\infty}\mathrm{d}v\int_{|1-v|}^{1+v}\mathrm{d}u~I(u,v){P}_{\phi}(vk){P}_{\phi}(uk),\quad\quad
\n
where $\Omega_{r}$ is the radiation density parameter at present, and ${P}_{\phi}(k)$ is the power spectrum of $\phi$. Here $u$ and $v$ are two dimensionless variables and we neglect the contribution of relativistic degrees of freedom in Eq.~(\ref{omegagw}). The function $I(u,v)$ comes from integrating the conformal time in the convolution of the source term, $ \<\mathcal{S}_{\vec{k}}(\te)\mathcal{S}_{\vec{k'}}(\te)\>$. In radiation dominated era, $I(u,v)$ is given by \cite{Espinosa:2018eve,Kohri:2018awv}
\m\label{i}
I(u,v)&&={27\over64}\(\frac{3(u^2+v^2-3)(-4v^2+(1-u^2+v^2)^2)}{16u^4v^4}\)^2\no\\
&&\Bigg(\Big(-4uv+(u^2+v^2-3)\log\bigg|\frac{3-(u+v)^2}{3-(u-v)^2}\bigg|\Big)^2\no\\
&&\qquad\qquad+\pi^2(u^2+v^2-3)^2\Theta(u+v-\sqrt{3})\Bigg),
\n
where $\Theta$ is the Heaviside function.

\begin{figure}[htbp!]
	\centering
	\includegraphics[width = 0.48\textwidth]{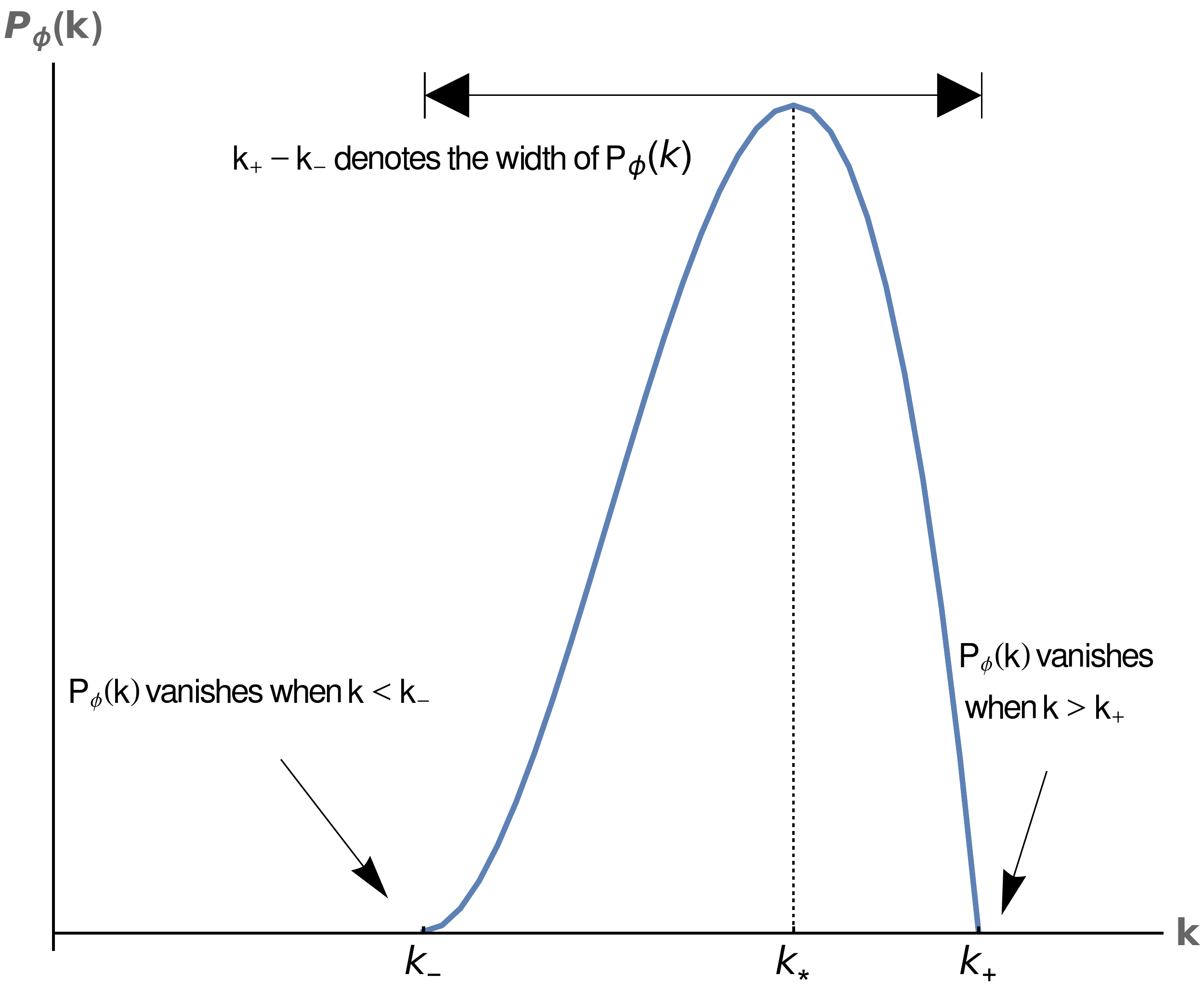}
	\caption{\label{pk} The schematic figure of the power spectrum $P_{\phi}(k)$ of scalar curvature perturbations. 
	}
\end{figure}

Let's consider a scalar power spectrum $P_\phi(k)$ which is peaked at $k_*$ and is nonzero only for $k_-<k<k_+$, like that illustrated in Fig.~\ref{pk}. 
Since the amplitude of the scalar power spectrum for the formation of PBHs is supposed to be much larger than those at CMB scales, the power spectrum at CMB scales is neglected.
For simplicity, we introduce a dimensionless parameter $\Delta$ to quantify the width of the scalar power spectrum as follows 
\m
\Delta\equiv {k_+-k_-\over k_*}. 
\n 
The power spectrum is narrow if $\Delta\ll 1$. 
From Eq.~(\ref{omegagw}), the density parameter of SIGWs for such a power spectrum reads 
\m
\Omega_{\mathrm{GW,eq}}
=\int_{k_-\over k}^{k_+\over k}\mathrm{d}v\int_{\max({k_-\over k},|1-v|)}^{\min({k_+\over k},1+v)}\mathrm{d}u I(u,v)P_\phi(uk)P_\phi(vk),\nonumber \\
\label{gaussian}
\n
In this letter, we focus on the behavior of $\Omega_{\mathrm{GW,eq}}(k)$ in the infrared region, namely $k\ll k_-$, and then $1\ll k_-/k<k_+/k$.

First of all, we consider a narrow power spectrum $(\Delta\ll 1)$, and divide the infrared region into two parts, i.e. $\Delta\ll k/k_*\ll 1$ and $k/k_*\ll \Delta$. For $\Delta\ll k/k_*\ll 1$, $k_-/k>v-1$ and $k_+/k<v+1$, and then 
\e
\Omega_{\mathrm{GW,eq}}(k)
=\int_{k_-\over k}^{k_+\over k}\mathrm{d}v\int_{k_-\over k}^{k_+\over k}\mathrm{d}uI(u,v)P_\phi(uk)P_\phi(vk). 
\q
Here $u,v \in [k_-/k, k_+/k] \gg 1$ and then 
\m
I(u,v)&&\simeq{243(u^2+v^2)^4(2(u^2+v^2)-(u^2-v^2)^2-1)^2\over 16384u^8v^8}\no\\
&&\times\ln^2{(u+v)^2\over 3}
\n
Introducing two new variables $(x, y)$ related to $(u, v)$ by $u={k_*\over k}(1+x)$ and $v={k_*\over k}(1+y)$ respectively, we have 
\m
\Omega_{\mathrm{GW,eq}}(k)
&&\simeq {243\over1024}\int_{-{k_*-k_-\over k_*}}^{k_+-k_*\over k_*}\mathrm{d}y\int_{-{k_*-k_-\over k_*}}^{k_+-k_*\over k_*}\mathrm{d}x~\no\\
&&\times {(1+x+y+{x^2\over 2}+{y^2\over 2})^4\over (1+x)^8 (1+y)^8} \no\\
&&\times\Big[(x-y)^2(2+x+y)^2\no\\
&&-4\({k\over k_*}\)^2(1+x+y+{x^2\over2}+{y^2\over2})\Big]^2\no\\
&&\times\ln^2 \[{4k_*^2\over 3k^2} (1+{x\over 2}+{y\over 2})^2\]\no \\
&&\times P_\phi(k_*(1+x))P_\phi(k_*(1+y)),\no\\
&&\simeq{243\over64}\({k\over k_*}\)^2\ln^2\({4k_*^2\over3k^2}\)P_\phi^2(k_*)\Delta^2.
\n
In the last step, we consider that both the absolute values of $x$ and $y$ are much smaller than one, and neglect the higher order corrections of order ${\cal O}(\Delta^3)$. And then the slope of SIGWs is given by 
\e\label{slope1}
n_{\mathrm{GW}}(k)=\frac{\mathrm{d}\log \Omega_{\mathrm{GW}}}{\mathrm{d}\log k}=2-{4\over\ln {4k_*^2\over3k^2}}.
\q
Switching to $k/k_*\ll \Delta$, we divide the integral range of $v$ in Eq.~(\ref{gaussian}) into three parts, namely
\m\label{intrange}
\Omega_{\mathrm{GW,eq}}(k)&&=\Bigg\{\int_{k_-\over k}^{{k_-\over k}+1}\mathrm{d}v\int_{k_-\over k}^{v+1}\mathrm{d}u+\int_{{k_-\over k}+1}^{{k_+\over k}-1}\mathrm{d}v\int_{v-1}^{v+1}\mathrm{d}u\no\\
&&+\int_{{k_+\over k}-1}^{k_+\over k}\mathrm{d}v\int_{v-1}^{k_+\over k}\mathrm{d}u\Bigg\}I(u,v)P_\phi(uk)P_\phi(vk).
\n
Considering $1\ll k_-/k<k_+/k$, both the first and the third integrations in the bracket of above equation are much small compared to the second integration which is approximately given by 
\m\label{cubic}
\Omega_{\mathrm{GW,eq}}(k)=2\int_{{k_-\over k}}^{{k_+\over k}}\mathrm{d}v I(v,v) P^2_\phi(vk). 
\n
Since $v\in [k_-/k, k_+/k]\gg 1$ and hence 
\m
I(v,v)\simeq {243\over 64}v^{-4} \ln^2{4v^2\over 3}, 
\n 
defining a new variable $y$ related to $v$ by $v={k_*\over k}(1+y)$, we have 
\m
\Omega_{\mathrm{GW,eq}}(k)&&\simeq {243\over32} \({k\over k_*}\)^3 \int_{-{k_*-k_-\over k_*}}^{{k_+-k_*\over k_*}}\mathrm{d}y (1+y)^{-4}\nonumber \\
&&\times \ln^2\[{4k_*^2\over 3k^2}(1+y)^2\] P^2_\phi(k_*(1+y)) \nonumber \\
&&\simeq {243\over32} \({k\over k_*}\)^3 \ln^2 \({4k_*^2\over 3k^2}\) P^2_\phi(k_*) \Delta, 
\n
where the higher order corrections of order $ {\cal O}(\Delta^2)$ are neglected as well. Similarly, the slope of SIGWs becomes 
\e\label{slope2}
n_{\mathrm{GW}}(k)=3-{4\over\ln {4k_*^2\over3k^2}}.
\q
In the infrared limit $(k\rightarrow 0)$, the slope approaches to $3$, or equivalently $\Omega_{\mathrm{GW}}(k\rightarrow 0)\propto k^3$, due to the un-correlation of the perturbations at those scales. In addition, for the $\delta$-power spectrum corresponding to $\Delta\rightarrow 0$, we only have $\Delta\ll k/k_*\ll 1$, and then $n_{\mathrm{GW}}=2-{4\over\ln {(4k_*^2/3k^2)}}$ which is consistent with that in \cite{Cai:2019cdl}.

From now on, we generalize our former discussion to the wide scalar power spectrum. In this case there is no available region of $\Delta\ll k/k_*\ll 1$ any more. Similar to the previous case of $k/k_*\ll \Delta$, one can easily find 
\m
\Omega_{\mathrm{GW,eq}}(k)&&={243\over32}\int_{{k_-\over k}}^{{k_+\over k}}\mathrm{d}v~v^{-4} \ln^2\(4v^2\over 3\) P^2_\phi(vk), \\
&&={243\over 32}k^3 \int_{k_-}^{k_+} dq~ q^{-4} \ln^2\(4q^2\over 3k^2\) P^2_\phi(q),~
\n
and then taking the derivative of $\log\Omega_{\mathrm{GW,eq}}$ with respect to $\log k$, we obtain 
\m\label{gs1}
n_{\mathrm{GW}}(k)&&=3-4\frac{\int_{k_-}^{k_+}\mathrm{d}p~p^{-4}\ln\(4p^2/3k^2\)P_\phi^2(p)}{\int_{k_-}^{k_+}\mathrm{d}q~q^{-4}\ln^2\(4q^2/3k^2\)P_\phi^2(q)}, \\
&&=3-4\frac{\int_{k_-/k}^{k_+/k}\mathrm{d}u~u^{-4}\ln\(4u^2/3\)P_\phi^2(uk)}{\int_{k_-/k}^{k_+/k}\mathrm{d}v~v^{-4}\ln^2\(4v^2/3\)P_\phi^2(vk)}. \label{ngwr}
\n
Strictly, $n_{\mathrm{GW}}$ given in the above equation certainly depends on the scalar power spectrum even for $k\ll k_-<k_+$. Notice that both $u$ and $v$ are much larger than one because $k_+/k>k_-/k\gg 1$ in the infrared region for wide scalar power spectrum, and both $\ln (4v^2/3)$ and $\ln(4u^2/3)$ can be roughly taken as a constant if $k_+$ is not larger than $k_-$ too much. In this sense, the above equation roughly gives  
\m
n_{\mathrm{GW}}(k)\approx 3-{4\over \ln {4k_\star^2\over 3k^2}}, 
\label{slope3}
\n
where $k_\star$ denotes a pivot scale for counting the integrations in Eq.~(\ref{ngwr}).

\begin{figure}[htbp!]
	\centering
	\includegraphics[width = 0.48\textwidth]{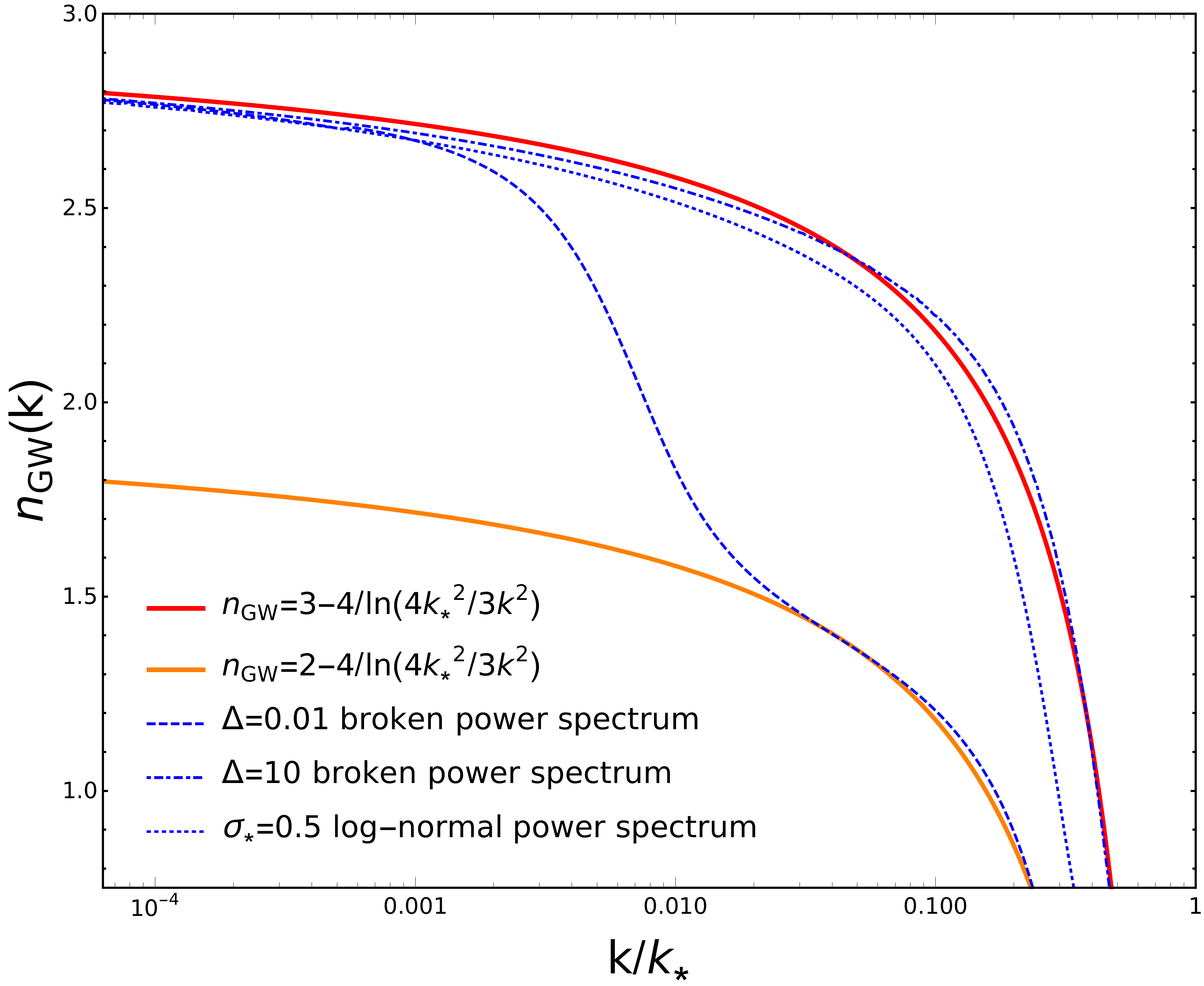}
	\caption{\label{gp} The slope of $\Omega_{\mathrm{GW}}(k)$. 
The orange and red solid lines are our analytic results in Eqs.~(\ref{slope1}) and (\ref{slope3}). The blue dashed and dot-dashed lines are the numerical results for the narrow $(k_-=0.995k_*, k_+=1.005k_*)$ and wide $(k_-=0.5, k_+=10.5 k_*)$ broken power spectra, and the blue dotted line is the numerical result for the log-normal power spectrum with $\sigma_*=0.5$, respectively.
	}
\end{figure}

Before closing this article, for example, we consider two power spectra: one is the broken power spectrum parameterized by
\begin{equation}\label{bps}
P_\phi(k)=\mathcal{A}\times \left\{
\begin{aligned}
{\frac{k-k_-}{k_*-k_-}}~&,&\ \hbox{for}\ k_-<k<k_*, \\
{\frac{k_+-k}{k_+-k_*}}~&,&\ \hbox{for}\ k_*<k<k_+, \\
\end{aligned}
\right.
\end{equation}
where $\mathcal{A}$ is the amplitude and $P_\phi(k)=0$ if $k<k_-$ or $k>k_+$; the other is a log-normal power spectrum, i.e.  
\e\label{ln}
P_\phi(k)={\mathcal{A}\over \sqrt{2\pi\sigma_*^2}}\exp\(-\frac{\ln(k/k_*)^2}{2\sigma_*^2}\)
\q
which has been studied in the literature (see e.g., \cite{Bartolo:2018rku}).
Here we consider one narrow power spectrum case ($\Delta=0.01$ in the broken power spectrum) and two wide power spectra ($\Delta=10$ in the broken power spectrum and $\sigma_*=0.5$ in the log-normal power spectrum) respectively. The analytic (orange and red solid lines) and numerical results (blue dashed, dotted and dot-dashed lines) are shown in Fig.~\ref{gp}.
For the narrow power spectrum ($\Delta=0.01$ in the broken power spectrum), the blue dashed line shows that $n_{\mathrm{GW}}(k)$ is roughly equal to $2-4/\ln(4k_*^2/3k^2)$ for $\Delta\ll k/k_*\ll 1$ and approaches to $3-4/\ln(4k_*^2/3k^2)$ for $k/k_*\ll \Delta$. For the wide power spectra ($\Delta=10$ in the broken power spectrum and $\sigma_*=0.5$ in the log-normal power spectrum), both the blue dot-dashed and blue dotted lines roughly recover our analytic result (the red line). Fig.~\ref{gp} indicates that our analytic results are nicely consistent with the numerical results.

To summarize, there are various GW sources, including the scalar curvature perturbation,  which generate a SGWB in the Universe. It is important to figure out some features of SIGW for distinguishing it from other sources. In this letter, we calculate the SIGW in the infrared region and find a log-dependent slope of SIGW, namely 
\m
n_{\mathrm{GW}}(f)=3-{2\over \ln(f_c/f)}, 
\n
and $n_{\mathrm{GW}}(f)=2-{2/ \ln(f_c/f)}$ near the peak if the scalar power spectrum is quite narrow. Here $f<f_c$ and $f_c$ is roughly the peak frequency. Even though the slope of SIGW approaches to 3 in the infrared limit, the correction of $2/\ln(f_c/f)$ approaches to zero slowly, and the amplitude of $\Omega_{\mathrm{GW}}(f)$ in the region corresponding to $\Omega_{\mathrm{GW}}(f)\propto f^3$ should be very small and could not be detected. In addition, the signal of gravitational waves usually is quite weak for the detectors and the template is generally needed for the gravitational-wave data analysis. Our results provide a new template for the data analysis. 



\begin{figure}[htbp!]
	\centering
	\includegraphics[width = 0.48\textwidth]{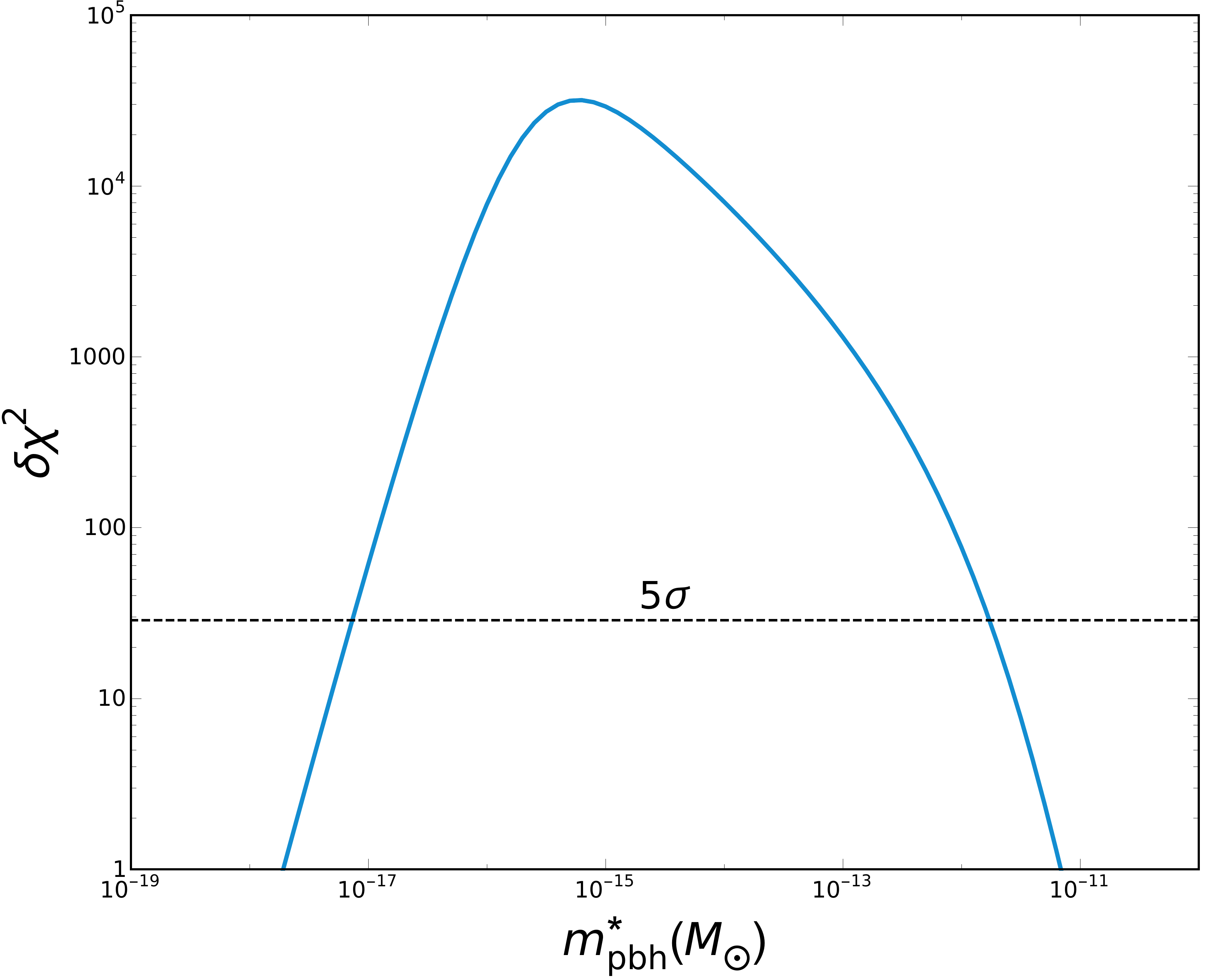}
	\caption{\label{fig:chi} The plot of $\delta\chi^2$ vs. the peak mass of PBHs ($\mpbh^*/M_\odot \approx2.3\times10^{18}\left({2\pi H_{0}}/{k_*}\right)^{2}$) generated by the log-normal power spectrum with $\sigma_*=0.5$. The amplitude $\mathcal{A}$ is fixed by assuming PBHs represent $10^{-3}$ of DM \cite{Yuan:2019udt}. The dashed line corresponds to $\delta\chi^2=28.74$. 
	}
\end{figure}

Furthermore, such a log-dependent slope of SIGWs is distinguishable for the future experiments, such as LISA \cite{Audley:2017drz}. For instance, we can compare our result of $\ogw$ for the log-normal power spectrum with $\sigma_*=0.5$ to a fiducial model with $n_{\mathrm{GW}}=3$ in the infrared region, namely
\begin{equation}
\ogw^{\rm{fid}}(k)=\left\{
\begin{aligned}
&\ogw(0.1k_*){\({k\over 0.1k_*}\)^3}~,&&\ \hbox{for}\ k<0.1k_*, \\
&\ogw(k)~,&&\ \hbox{for}\ k\ge0.1k_*. \\
\end{aligned}
\right.
\end{equation}
In order to give a quantitative estimation for the distinguishability, we need to calculate $\delta\chi^2$ given in \cite{Kuroyanagi:2018csn}, namely 
\m
\delta\chi^2\simeq T\int_0^\infty\mathrm{d}f\(\frac{\ogw-\ogw^{\rm{fid}}}{\ogw^{\rm{fid}}+\Omega_{n}}\)^2,
\n
where $\Om_n(f) = 2\pi^2 f^3 S_n/(3H_0^2)$
and $S_n$ is the strain power spectral density of LISA \cite{Cornish:2018dyw} and the observation time of LISA is set to be $T=1$yr typically. In principle, $\delta\chi^2>28.74$ implies that we can distinguishing our result from the fiducial model at more than $5\sigma$ confidence level. The result of $\chi^2$ in Fig.~\ref{fig:chi} indicates that LISA can well distinguish our prediction for SIGWs from the fiducial model at high confidence level.

Actually, such a log-dependent slope is quite generic for SIGW. It comes from the oscillating behavior of the evolution of the scalar perturbations (the sine and cosine terms in Eq.~(\ref{phi})). Integrating over these sine and cosine terms results in a cosine integral, $\mathrm{Ci}(x)$, and 
\e
\lim_{x\rightarrow0^+} \mathrm{Ci}(|A|x)-\mathrm{Ci}(|B|x)=\ln(|A/B|),
\q
which finally leads to the logarithmic terms in $I(u,v)$. In this sense, the log-dependent slope of the SIGW spectra is a result of the evolution of the scalar perturbations in a radiation dominated universe. It implies that such a log-dependent slope is a unique feature for the SIGW which can be used to distinguish SIGW from other sources.

Acknowledgments. We acknowledge the use of HPC Cluster of ITP-CAS. 
	This work is supported by grants from NSFC 
	(grant No. 11690021, 11975019, 11947302, 11991053), 
	the Strategic Priority Research Program of Chinese Academy of Sciences 
	(Grant No. XDB23000000, XDA15020701), and Key Research Program of Frontier Sciences, CAS, Grant NO. ZDBS-LY-7009.

	\bibliography{./ngref}

\begin{thebibliography}{73}%
\makeatletter
\providecommand \@ifxundefined [1]{%
 \@ifx{#1\undefined}
}%
\providecommand \@ifnum [1]{%
 \ifnum #1\expandafter \@firstoftwo
 \else \expandafter \@secondoftwo
 \fi
}%
\providecommand \@ifx [1]{%
 \ifx #1\expandafter \@firstoftwo
 \else \expandafter \@secondoftwo
 \fi
}%
\providecommand \natexlab [1]{#1}%
\providecommand \enquote  [1]{``#1''}%
\providecommand \bibnamefont  [1]{#1}%
\providecommand \bibfnamefont [1]{#1}%
\providecommand \citenamefont [1]{#1}%
\providecommand \href@noop [0]{\@secondoftwo}%
\providecommand \href [0]{\begingroup \@sanitize@url \@href}%
\providecommand \@href[1]{\@@startlink{#1}\@@href}%
\providecommand \@@href[1]{\endgroup#1\@@endlink}%
\providecommand \@sanitize@url [0]{\catcode `\\12\catcode `\$12\catcode
  `\&12\catcode `\#12\catcode `\^12\catcode `\_12\catcode `\%12\relax}%
\providecommand \@@startlink[1]{}%
\providecommand \@@endlink[0]{}%
\providecommand \url  [0]{\begingroup\@sanitize@url \@url }%
\providecommand \@url [1]{\endgroup\@href {#1}{\urlprefix }}%
\providecommand \urlprefix  [0]{URL }%
\providecommand \Eprint [0]{\href }%
\providecommand \doibase [0]{http://dx.doi.org/}%
\providecommand \selectlanguage [0]{\@gobble}%
\providecommand \bibinfo  [0]{\@secondoftwo}%
\providecommand \bibfield  [0]{\@secondoftwo}%
\providecommand \translation [1]{[#1]}%
\providecommand \BibitemOpen [0]{}%
\providecommand \bibitemStop [0]{}%
\providecommand \bibitemNoStop [0]{.\EOS\space}%
\providecommand \EOS [0]{\spacefactor3000\relax}%
\providecommand \BibitemShut  [1]{\csname bibitem#1\endcsname}%
\let\auto@bib@innerbib\@empty
\bibitem [{\citenamefont {Aghanim}\ \emph {et~al.}(2018)\citenamefont {Aghanim}
  \emph {et~al.}}]{Aghanim:2018eyx}%
  \BibitemOpen
  \bibfield  {author} {\bibinfo {author} {\bibfnamefont {N.}~\bibnamefont
  {Aghanim}} \emph {et~al.} (\bibinfo {collaboration} {Planck}),\ }\href@noop
  {} {\  (\bibinfo {year} {2018})},\ \Eprint {http://arxiv.org/abs/1807.06209}
  {arXiv:1807.06209 [astro-ph.CO]} \BibitemShut {NoStop}%
\bibitem [{\citenamefont {Abbott}\ \emph {et~al.}(2016)\citenamefont {Abbott}
  \emph {et~al.}}]{Abbott:2016blz}%
  \BibitemOpen
  \bibfield  {author} {\bibinfo {author} {\bibfnamefont {B.~P.}\ \bibnamefont
  {Abbott}} \emph {et~al.} (\bibinfo {collaboration} {LIGO Scientific,
  Virgo}),\ }\href {\doibase 10.1103/PhysRevLett.116.061102} {\bibfield
  {journal} {\bibinfo  {journal} {Phys. Rev. Lett.}\ }\textbf {\bibinfo
  {volume} {116}},\ \bibinfo {pages} {061102} (\bibinfo {year} {2016})},\
  \Eprint {http://arxiv.org/abs/1602.03837} {arXiv:1602.03837 [gr-qc]}
  \BibitemShut {NoStop}%
\bibitem [{\citenamefont {Sasaki}\ \emph {et~al.}(2016)\citenamefont {Sasaki},
  \citenamefont {Suyama}, \citenamefont {Tanaka},\ and\ \citenamefont
  {Yokoyama}}]{Sasaki:2016jop}%
  \BibitemOpen
  \bibfield  {author} {\bibinfo {author} {\bibfnamefont {M.}~\bibnamefont
  {Sasaki}}, \bibinfo {author} {\bibfnamefont {T.}~\bibnamefont {Suyama}},
  \bibinfo {author} {\bibfnamefont {T.}~\bibnamefont {Tanaka}}, \ and\ \bibinfo
  {author} {\bibfnamefont {S.}~\bibnamefont {Yokoyama}},\ }\href {\doibase
  10.1103/PhysRevLett.121.059901, 10.1103/PhysRevLett.117.061101} {\bibfield
  {journal} {\bibinfo  {journal} {Phys. Rev. Lett.}\ }\textbf {\bibinfo
  {volume} {117}},\ \bibinfo {pages} {061101} (\bibinfo {year} {2016})},\
  \bibinfo {note} {[erratum: Phys. Rev. Lett.121,no.5,059901(2018)]},\ \Eprint
  {http://arxiv.org/abs/1603.08338} {arXiv:1603.08338 [astro-ph.CO]}
  \BibitemShut {NoStop}%
\bibitem [{\citenamefont {Chen}\ and\ \citenamefont
  {Huang}(2018)}]{Chen:2018czv}%
  \BibitemOpen
  \bibfield  {author} {\bibinfo {author} {\bibfnamefont {Z.-C.}\ \bibnamefont
  {Chen}}\ and\ \bibinfo {author} {\bibfnamefont {Q.-G.}\ \bibnamefont
  {Huang}},\ }\href {\doibase 10.3847/1538-4357/aad6e2} {\bibfield  {journal}
  {\bibinfo  {journal} {Astrophys. J.}\ }\textbf {\bibinfo {volume} {864}},\
  \bibinfo {pages} {61} (\bibinfo {year} {2018})},\ \Eprint
  {http://arxiv.org/abs/1801.10327} {arXiv:1801.10327 [astro-ph.CO]}
  \BibitemShut {NoStop}%
\bibitem [{\citenamefont {Chen}\ and\ \citenamefont
  {Huang}(2019)}]{Chen:2019irf}%
  \BibitemOpen
  \bibfield  {author} {\bibinfo {author} {\bibfnamefont {Z.-C.}\ \bibnamefont
  {Chen}}\ and\ \bibinfo {author} {\bibfnamefont {Q.-G.}\ \bibnamefont
  {Huang}},\ }\href@noop {} {\  (\bibinfo {year} {2019})},\ \Eprint
  {http://arxiv.org/abs/1904.02396} {arXiv:1904.02396 [astro-ph.CO]}
  \BibitemShut {NoStop}%
\bibitem [{\citenamefont {Chen}\ \emph
  {et~al.}(2019{\natexlab{a}})\citenamefont {Chen}, \citenamefont {Huang},\
  and\ \citenamefont {Huang}}]{Chen:2018rzo}%
  \BibitemOpen
  \bibfield  {author} {\bibinfo {author} {\bibfnamefont {Z.-C.}\ \bibnamefont
  {Chen}}, \bibinfo {author} {\bibfnamefont {F.}~\bibnamefont {Huang}}, \ and\
  \bibinfo {author} {\bibfnamefont {Q.-G.}\ \bibnamefont {Huang}},\ }\href
  {\doibase 10.3847/1538-4357/aaf581} {\bibfield  {journal} {\bibinfo
  {journal} {Astrophys. J.}\ }\textbf {\bibinfo {volume} {871}},\ \bibinfo
  {pages} {97} (\bibinfo {year} {2019}{\natexlab{a}})},\ \Eprint
  {http://arxiv.org/abs/1809.10360} {arXiv:1809.10360 [gr-qc]} \BibitemShut
  {NoStop}%
\bibitem [{\citenamefont {Tisserand}\ \emph {et~al.}(2007)\citenamefont
  {Tisserand} \emph {et~al.}}]{Tisserand:2006zx}%
  \BibitemOpen
  \bibfield  {author} {\bibinfo {author} {\bibfnamefont {P.}~\bibnamefont
  {Tisserand}} \emph {et~al.} (\bibinfo {collaboration} {EROS-2}),\ }\href
  {\doibase 10.1051/0004-6361:20066017} {\bibfield  {journal} {\bibinfo
  {journal} {Astron. Astrophys.}\ }\textbf {\bibinfo {volume} {469}},\ \bibinfo
  {pages} {387} (\bibinfo {year} {2007})},\ \Eprint
  {http://arxiv.org/abs/astro-ph/0607207} {arXiv:astro-ph/0607207 [astro-ph]}
  \BibitemShut {NoStop}%
\bibitem [{\citenamefont {Carr}\ \emph {et~al.}(2010)\citenamefont {Carr},
  \citenamefont {Kohri}, \citenamefont {Sendouda},\ and\ \citenamefont
  {Yokoyama}}]{Carr:2009jm}%
  \BibitemOpen
  \bibfield  {author} {\bibinfo {author} {\bibfnamefont {B.~J.}\ \bibnamefont
  {Carr}}, \bibinfo {author} {\bibfnamefont {K.}~\bibnamefont {Kohri}},
  \bibinfo {author} {\bibfnamefont {Y.}~\bibnamefont {Sendouda}}, \ and\
  \bibinfo {author} {\bibfnamefont {J.}~\bibnamefont {Yokoyama}},\ }\href
  {\doibase 10.1103/PhysRevD.81.104019} {\bibfield  {journal} {\bibinfo
  {journal} {Phys. Rev.}\ }\textbf {\bibinfo {volume} {D81}},\ \bibinfo {pages}
  {104019} (\bibinfo {year} {2010})},\ \Eprint {http://arxiv.org/abs/0912.5297}
  {arXiv:0912.5297 [astro-ph.CO]} \BibitemShut {NoStop}%
\bibitem [{\citenamefont {Barnacka}\ \emph {et~al.}(2012)\citenamefont
  {Barnacka}, \citenamefont {Glicenstein},\ and\ \citenamefont
  {Moderski}}]{Barnacka:2012bm}%
  \BibitemOpen
  \bibfield  {author} {\bibinfo {author} {\bibfnamefont {A.}~\bibnamefont
  {Barnacka}}, \bibinfo {author} {\bibfnamefont {J.~F.}\ \bibnamefont
  {Glicenstein}}, \ and\ \bibinfo {author} {\bibfnamefont {R.}~\bibnamefont
  {Moderski}},\ }\href {\doibase 10.1103/PhysRevD.86.043001} {\bibfield
  {journal} {\bibinfo  {journal} {Phys. Rev.}\ }\textbf {\bibinfo {volume}
  {D86}},\ \bibinfo {pages} {043001} (\bibinfo {year} {2012})},\ \Eprint
  {http://arxiv.org/abs/1204.2056} {arXiv:1204.2056 [astro-ph.CO]} \BibitemShut
  {NoStop}%
\bibitem [{\citenamefont {Griest}\ \emph {et~al.}(2013)\citenamefont {Griest},
  \citenamefont {Cieplak},\ and\ \citenamefont {Lehner}}]{Griest:2013esa}%
  \BibitemOpen
  \bibfield  {author} {\bibinfo {author} {\bibfnamefont {K.}~\bibnamefont
  {Griest}}, \bibinfo {author} {\bibfnamefont {A.~M.}\ \bibnamefont {Cieplak}},
  \ and\ \bibinfo {author} {\bibfnamefont {M.~J.}\ \bibnamefont {Lehner}},\
  }\href {\doibase 10.1103/PhysRevLett.111.181302} {\bibfield  {journal}
  {\bibinfo  {journal} {Phys. Rev. Lett.}\ }\textbf {\bibinfo {volume} {111}},\
  \bibinfo {pages} {181302} (\bibinfo {year} {2013})}\BibitemShut {NoStop}%
\bibitem [{\citenamefont {Graham}\ \emph {et~al.}(2015)\citenamefont {Graham},
  \citenamefont {Rajendran},\ and\ \citenamefont {Varela}}]{Graham:2015apa}%
  \BibitemOpen
  \bibfield  {author} {\bibinfo {author} {\bibfnamefont {P.~W.}\ \bibnamefont
  {Graham}}, \bibinfo {author} {\bibfnamefont {S.}~\bibnamefont {Rajendran}}, \
  and\ \bibinfo {author} {\bibfnamefont {J.}~\bibnamefont {Varela}},\ }\href
  {\doibase 10.1103/PhysRevD.92.063007} {\bibfield  {journal} {\bibinfo
  {journal} {Phys. Rev.}\ }\textbf {\bibinfo {volume} {D92}},\ \bibinfo {pages}
  {063007} (\bibinfo {year} {2015})},\ \Eprint
  {http://arxiv.org/abs/1505.04444} {arXiv:1505.04444 [hep-ph]} \BibitemShut
  {NoStop}%
\bibitem [{\citenamefont {Brandt}(2016)}]{Brandt:2016aco}%
  \BibitemOpen
  \bibfield  {author} {\bibinfo {author} {\bibfnamefont {T.~D.}\ \bibnamefont
  {Brandt}},\ }\href {\doibase 10.3847/2041-8205/824/2/L31} {\bibfield
  {journal} {\bibinfo  {journal} {Astrophys. J.}\ }\textbf {\bibinfo {volume}
  {824}},\ \bibinfo {pages} {L31} (\bibinfo {year} {2016})},\ \Eprint
  {http://arxiv.org/abs/1605.03665} {arXiv:1605.03665 [astro-ph.GA]}
  \BibitemShut {NoStop}%
\bibitem [{\citenamefont {Chen}\ \emph {et~al.}(2016)\citenamefont {Chen},
  \citenamefont {Huang},\ and\ \citenamefont {Wang}}]{Chen:2016pud}%
  \BibitemOpen
  \bibfield  {author} {\bibinfo {author} {\bibfnamefont {L.}~\bibnamefont
  {Chen}}, \bibinfo {author} {\bibfnamefont {Q.-G.}\ \bibnamefont {Huang}}, \
  and\ \bibinfo {author} {\bibfnamefont {K.}~\bibnamefont {Wang}},\ }\href
  {\doibase 10.1088/1475-7516/2016/12/044} {\bibfield  {journal} {\bibinfo
  {journal} {JCAP}\ }\textbf {\bibinfo {volume} {1612}},\ \bibinfo {pages}
  {044} (\bibinfo {year} {2016})},\ \Eprint {http://arxiv.org/abs/1608.02174}
  {arXiv:1608.02174 [astro-ph.CO]} \BibitemShut {NoStop}%
\bibitem [{\citenamefont {Wang}\ \emph {et~al.}(2018)\citenamefont {Wang},
  \citenamefont {Wang}, \citenamefont {Huang},\ and\ \citenamefont
  {Li}}]{Wang:2016ana}%
  \BibitemOpen
  \bibfield  {author} {\bibinfo {author} {\bibfnamefont {S.}~\bibnamefont
  {Wang}}, \bibinfo {author} {\bibfnamefont {Y.-F.}\ \bibnamefont {Wang}},
  \bibinfo {author} {\bibfnamefont {Q.-G.}\ \bibnamefont {Huang}}, \ and\
  \bibinfo {author} {\bibfnamefont {T.~G.~F.}\ \bibnamefont {Li}},\ }\href
  {\doibase 10.1103/PhysRevLett.120.191102} {\bibfield  {journal} {\bibinfo
  {journal} {Phys. Rev. Lett.}\ }\textbf {\bibinfo {volume} {120}},\ \bibinfo
  {pages} {191102} (\bibinfo {year} {2018})},\ \Eprint
  {http://arxiv.org/abs/1610.08725} {arXiv:1610.08725 [astro-ph.CO]}
  \BibitemShut {NoStop}%
\bibitem [{\citenamefont {Gaggero}\ \emph {et~al.}(2017)\citenamefont
  {Gaggero}, \citenamefont {Bertone}, \citenamefont {Calore}, \citenamefont
  {Connors}, \citenamefont {Lovell}, \citenamefont {Markoff},\ and\
  \citenamefont {Storm}}]{Gaggero:2016dpq}%
  \BibitemOpen
  \bibfield  {author} {\bibinfo {author} {\bibfnamefont {D.}~\bibnamefont
  {Gaggero}}, \bibinfo {author} {\bibfnamefont {G.}~\bibnamefont {Bertone}},
  \bibinfo {author} {\bibfnamefont {F.}~\bibnamefont {Calore}}, \bibinfo
  {author} {\bibfnamefont {R.~M.~T.}\ \bibnamefont {Connors}}, \bibinfo
  {author} {\bibfnamefont {M.}~\bibnamefont {Lovell}}, \bibinfo {author}
  {\bibfnamefont {S.}~\bibnamefont {Markoff}}, \ and\ \bibinfo {author}
  {\bibfnamefont {E.}~\bibnamefont {Storm}},\ }\href {\doibase
  10.1103/PhysRevLett.118.241101} {\bibfield  {journal} {\bibinfo  {journal}
  {Phys. Rev. Lett.}\ }\textbf {\bibinfo {volume} {118}},\ \bibinfo {pages}
  {241101} (\bibinfo {year} {2017})},\ \Eprint
  {http://arxiv.org/abs/1612.00457} {arXiv:1612.00457 [astro-ph.HE]}
  \BibitemShut {NoStop}%
\bibitem [{\citenamefont {Ali-Haimoud}\ and\ \citenamefont
  {Kamionkowski}(2017)}]{Ali-Haimoud:2016mbv}%
  \BibitemOpen
  \bibfield  {author} {\bibinfo {author} {\bibfnamefont {Y.}~\bibnamefont
  {Ali-Haimoud}}\ and\ \bibinfo {author} {\bibfnamefont {M.}~\bibnamefont
  {Kamionkowski}},\ }\href {\doibase 10.1103/PhysRevD.95.043534} {\bibfield
  {journal} {\bibinfo  {journal} {Phys. Rev.}\ }\textbf {\bibinfo {volume}
  {D95}},\ \bibinfo {pages} {043534} (\bibinfo {year} {2017})},\ \Eprint
  {http://arxiv.org/abs/1612.05644} {arXiv:1612.05644 [astro-ph.CO]}
  \BibitemShut {NoStop}%
\bibitem [{\citenamefont {Aloni}\ \emph {et~al.}(2017)\citenamefont {Aloni},
  \citenamefont {Blum},\ and\ \citenamefont {Flauger}}]{Blum:2016cjs}%
  \BibitemOpen
  \bibfield  {author} {\bibinfo {author} {\bibfnamefont {D.}~\bibnamefont
  {Aloni}}, \bibinfo {author} {\bibfnamefont {K.}~\bibnamefont {Blum}}, \ and\
  \bibinfo {author} {\bibfnamefont {R.}~\bibnamefont {Flauger}},\ }\href
  {\doibase 10.1088/1475-7516/2017/05/017} {\bibfield  {journal} {\bibinfo
  {journal} {JCAP}\ }\textbf {\bibinfo {volume} {1705}},\ \bibinfo {pages}
  {017} (\bibinfo {year} {2017})},\ \Eprint {http://arxiv.org/abs/1612.06811}
  {arXiv:1612.06811 [astro-ph.CO]} \BibitemShut {NoStop}%
\bibitem [{\citenamefont {Horowitz}(2016)}]{Horowitz:2016lib}%
  \BibitemOpen
  \bibfield  {author} {\bibinfo {author} {\bibfnamefont {B.}~\bibnamefont
  {Horowitz}},\ }\href@noop {} {\  (\bibinfo {year} {2016})},\ \Eprint
  {http://arxiv.org/abs/1612.07264} {arXiv:1612.07264 [astro-ph.CO]}
  \BibitemShut {NoStop}%
\bibitem [{\citenamefont {Niikura}\ \emph
  {et~al.}(2019{\natexlab{a}})\citenamefont {Niikura} \emph
  {et~al.}}]{Niikura:2017zjd}%
  \BibitemOpen
  \bibfield  {author} {\bibinfo {author} {\bibfnamefont {H.}~\bibnamefont
  {Niikura}} \emph {et~al.},\ }\href {\doibase 10.1038/s41550-019-0723-1}
  {\bibfield  {journal} {\bibinfo  {journal} {Nat. Astron.}\ }\textbf {\bibinfo
  {volume} {3}},\ \bibinfo {pages} {524} (\bibinfo {year}
  {2019}{\natexlab{a}})},\ \Eprint {http://arxiv.org/abs/1701.02151}
  {arXiv:1701.02151 [astro-ph.CO]} \BibitemShut {NoStop}%
\bibitem [{\citenamefont {Zumalacarregui}\ and\ \citenamefont
  {Seljak}(2018)}]{Zumalacarregui:2017qqd}%
  \BibitemOpen
  \bibfield  {author} {\bibinfo {author} {\bibfnamefont {M.}~\bibnamefont
  {Zumalacarregui}}\ and\ \bibinfo {author} {\bibfnamefont {U.}~\bibnamefont
  {Seljak}},\ }\href {\doibase 10.1103/PhysRevLett.121.141101} {\bibfield
  {journal} {\bibinfo  {journal} {Phys. Rev. Lett.}\ }\textbf {\bibinfo
  {volume} {121}},\ \bibinfo {pages} {141101} (\bibinfo {year} {2018})},\
  \Eprint {http://arxiv.org/abs/1712.02240} {arXiv:1712.02240 [astro-ph.CO]}
  \BibitemShut {NoStop}%
\bibitem [{\citenamefont {Abbott}\ \emph {et~al.}(2018)\citenamefont {Abbott}
  \emph {et~al.}}]{Abbott:2018oah}%
  \BibitemOpen
  \bibfield  {author} {\bibinfo {author} {\bibfnamefont {B.~P.}\ \bibnamefont
  {Abbott}} \emph {et~al.} (\bibinfo {collaboration} {LIGO Scientific,
  Virgo}),\ }\href {\doibase 10.1103/PhysRevLett.121.231103} {\bibfield
  {journal} {\bibinfo  {journal} {Phys. Rev. Lett.}\ }\textbf {\bibinfo
  {volume} {121}},\ \bibinfo {pages} {231103} (\bibinfo {year} {2018})},\
  \Eprint {http://arxiv.org/abs/1808.04771} {arXiv:1808.04771 [astro-ph.CO]}
  \BibitemShut {NoStop}%
\bibitem [{\citenamefont {Magee}\ \emph {et~al.}(2018)\citenamefont {Magee},
  \citenamefont {Deutsch}, \citenamefont {McClincy}, \citenamefont {Hanna},
  \citenamefont {Horst}, \citenamefont {Meacher}, \citenamefont {Messick},
  \citenamefont {Shandera},\ and\ \citenamefont {Wade}}]{Magee:2018opb}%
  \BibitemOpen
  \bibfield  {author} {\bibinfo {author} {\bibfnamefont {R.}~\bibnamefont
  {Magee}}, \bibinfo {author} {\bibfnamefont {A.-S.}\ \bibnamefont {Deutsch}},
  \bibinfo {author} {\bibfnamefont {P.}~\bibnamefont {McClincy}}, \bibinfo
  {author} {\bibfnamefont {C.}~\bibnamefont {Hanna}}, \bibinfo {author}
  {\bibfnamefont {C.}~\bibnamefont {Horst}}, \bibinfo {author} {\bibfnamefont
  {D.}~\bibnamefont {Meacher}}, \bibinfo {author} {\bibfnamefont
  {C.}~\bibnamefont {Messick}}, \bibinfo {author} {\bibfnamefont
  {S.}~\bibnamefont {Shandera}}, \ and\ \bibinfo {author} {\bibfnamefont
  {M.}~\bibnamefont {Wade}},\ }\href {\doibase 10.1103/PhysRevD.98.103024}
  {\bibfield  {journal} {\bibinfo  {journal} {Phys. Rev.}\ }\textbf {\bibinfo
  {volume} {D98}},\ \bibinfo {pages} {103024} (\bibinfo {year} {2018})},\
  \Eprint {http://arxiv.org/abs/1808.04772} {arXiv:1808.04772 [astro-ph.IM]}
  \BibitemShut {NoStop}%
\bibitem [{\citenamefont {Niikura}\ \emph
  {et~al.}(2019{\natexlab{b}})\citenamefont {Niikura}, \citenamefont {Takada},
  \citenamefont {Yokoyama}, \citenamefont {Sumi},\ and\ \citenamefont
  {Masaki}}]{Niikura:2019kqi}%
  \BibitemOpen
  \bibfield  {author} {\bibinfo {author} {\bibfnamefont {H.}~\bibnamefont
  {Niikura}}, \bibinfo {author} {\bibfnamefont {M.}~\bibnamefont {Takada}},
  \bibinfo {author} {\bibfnamefont {S.}~\bibnamefont {Yokoyama}}, \bibinfo
  {author} {\bibfnamefont {T.}~\bibnamefont {Sumi}}, \ and\ \bibinfo {author}
  {\bibfnamefont {S.}~\bibnamefont {Masaki}},\ }\href {\doibase
  10.1103/PhysRevD.99.083503} {\bibfield  {journal} {\bibinfo  {journal} {Phys.
  Rev.}\ }\textbf {\bibinfo {volume} {D99}},\ \bibinfo {pages} {083503}
  (\bibinfo {year} {2019}{\natexlab{b}})},\ \Eprint
  {http://arxiv.org/abs/1901.07120} {arXiv:1901.07120 [astro-ph.CO]}
  \BibitemShut {NoStop}%
\bibitem [{\citenamefont {Abbott}\ \emph {et~al.}(2019)\citenamefont {Abbott}
  \emph {et~al.}}]{Authors:2019qbw}%
  \BibitemOpen
  \bibfield  {author} {\bibinfo {author} {\bibfnamefont {B.~P.}\ \bibnamefont
  {Abbott}} \emph {et~al.} (\bibinfo {collaboration} {LIGO Scientific,
  Virgo}),\ }\href@noop {} {\  (\bibinfo {year} {2019})},\ \Eprint
  {http://arxiv.org/abs/1904.08976} {arXiv:1904.08976 [astro-ph.CO]}
  \BibitemShut {NoStop}%
\bibitem [{\citenamefont {Wang}\ \emph {et~al.}(2019)\citenamefont {Wang},
  \citenamefont {Huang}, \citenamefont {Li},\ and\ \citenamefont
  {Liao}}]{Wang:2019kzb}%
  \BibitemOpen
  \bibfield  {author} {\bibinfo {author} {\bibfnamefont {Y.-F.}\ \bibnamefont
  {Wang}}, \bibinfo {author} {\bibfnamefont {Q.-G.}\ \bibnamefont {Huang}},
  \bibinfo {author} {\bibfnamefont {T.~G.~F.}\ \bibnamefont {Li}}, \ and\
  \bibinfo {author} {\bibfnamefont {S.}~\bibnamefont {Liao}},\ }\href@noop {}
  {\  (\bibinfo {year} {2019})},\ \Eprint {http://arxiv.org/abs/1910.07397}
  {arXiv:1910.07397 [astro-ph.CO]} \BibitemShut {NoStop}%
\bibitem [{\citenamefont {Wu}(2020)}]{Wu:2020drm}%
  \BibitemOpen
  \bibfield  {author} {\bibinfo {author} {\bibfnamefont {Y.}~\bibnamefont
  {Wu}},\ }\href@noop {} {\  (\bibinfo {year} {2020})},\ \Eprint
  {http://arxiv.org/abs/2001.03833} {arXiv:2001.03833 [astro-ph.CO]}
  \BibitemShut {NoStop}%
\bibitem [{\citenamefont {Hawking}(1971)}]{Hawking:1971ei}%
  \BibitemOpen
  \bibfield  {author} {\bibinfo {author} {\bibfnamefont {S.}~\bibnamefont
  {Hawking}},\ }\href@noop {} {\bibfield  {journal} {\bibinfo  {journal} {Mon.
  Not. Roy. Astron. Soc.}\ }\textbf {\bibinfo {volume} {152}},\ \bibinfo
  {pages} {75} (\bibinfo {year} {1971})}\BibitemShut {NoStop}%
\bibitem [{\citenamefont {Carr}\ and\ \citenamefont
  {Hawking}(1974)}]{Carr:1974nx}%
  \BibitemOpen
  \bibfield  {author} {\bibinfo {author} {\bibfnamefont {B.~J.}\ \bibnamefont
  {Carr}}\ and\ \bibinfo {author} {\bibfnamefont {S.~W.}\ \bibnamefont
  {Hawking}},\ }\href@noop {} {\bibfield  {journal} {\bibinfo  {journal} {Mon.
  Not. Roy. Astron. Soc.}\ }\textbf {\bibinfo {volume} {168}},\ \bibinfo
  {pages} {399} (\bibinfo {year} {1974})}\BibitemShut {NoStop}%
\bibitem [{\citenamefont {Tomita}(1967)}]{tomita1967non}%
  \BibitemOpen
  \bibfield  {author} {\bibinfo {author} {\bibfnamefont {K.}~\bibnamefont
  {Tomita}},\ }\href@noop {} {\bibfield  {journal} {\bibinfo  {journal}
  {Progress of Theoretical Physics}\ }\textbf {\bibinfo {volume} {37}},\
  \bibinfo {pages} {831} (\bibinfo {year} {1967})}\BibitemShut {NoStop}%
\bibitem [{\citenamefont {Matarrese}\ \emph {et~al.}(1993)\citenamefont
  {Matarrese}, \citenamefont {Pantano},\ and\ \citenamefont
  {Saez}}]{Matarrese:1992rp}%
  \BibitemOpen
  \bibfield  {author} {\bibinfo {author} {\bibfnamefont {S.}~\bibnamefont
  {Matarrese}}, \bibinfo {author} {\bibfnamefont {O.}~\bibnamefont {Pantano}},
  \ and\ \bibinfo {author} {\bibfnamefont {D.}~\bibnamefont {Saez}},\ }\href
  {\doibase 10.1103/PhysRevD.47.1311} {\bibfield  {journal} {\bibinfo
  {journal} {Phys. Rev.}\ }\textbf {\bibinfo {volume} {D47}},\ \bibinfo {pages}
  {1311} (\bibinfo {year} {1993})}\BibitemShut {NoStop}%
\bibitem [{\citenamefont {Matarrese}\ \emph {et~al.}(1994)\citenamefont
  {Matarrese}, \citenamefont {Pantano},\ and\ \citenamefont
  {Saez}}]{Matarrese:1993zf}%
  \BibitemOpen
  \bibfield  {author} {\bibinfo {author} {\bibfnamefont {S.}~\bibnamefont
  {Matarrese}}, \bibinfo {author} {\bibfnamefont {O.}~\bibnamefont {Pantano}},
  \ and\ \bibinfo {author} {\bibfnamefont {D.}~\bibnamefont {Saez}},\ }\href
  {\doibase 10.1103/PhysRevLett.72.320} {\bibfield  {journal} {\bibinfo
  {journal} {Phys. Rev. Lett.}\ }\textbf {\bibinfo {volume} {72}},\ \bibinfo
  {pages} {320} (\bibinfo {year} {1994})},\ \Eprint
  {http://arxiv.org/abs/astro-ph/9310036} {arXiv:astro-ph/9310036 [astro-ph]}
  \BibitemShut {NoStop}%
\bibitem [{\citenamefont {Matarrese}\ \emph {et~al.}(1998)\citenamefont
  {Matarrese}, \citenamefont {Mollerach},\ and\ \citenamefont
  {Bruni}}]{Matarrese:1997ay}%
  \BibitemOpen
  \bibfield  {author} {\bibinfo {author} {\bibfnamefont {S.}~\bibnamefont
  {Matarrese}}, \bibinfo {author} {\bibfnamefont {S.}~\bibnamefont
  {Mollerach}}, \ and\ \bibinfo {author} {\bibfnamefont {M.}~\bibnamefont
  {Bruni}},\ }\href {\doibase 10.1103/PhysRevD.58.043504} {\bibfield  {journal}
  {\bibinfo  {journal} {Phys. Rev.}\ }\textbf {\bibinfo {volume} {D58}},\
  \bibinfo {pages} {043504} (\bibinfo {year} {1998})},\ \Eprint
  {http://arxiv.org/abs/astro-ph/9707278} {arXiv:astro-ph/9707278 [astro-ph]}
  \BibitemShut {NoStop}%
\bibitem [{\citenamefont {Noh}\ and\ \citenamefont {Hwang}(2004)}]{Noh:2004bc}%
  \BibitemOpen
  \bibfield  {author} {\bibinfo {author} {\bibfnamefont {H.}~\bibnamefont
  {Noh}}\ and\ \bibinfo {author} {\bibfnamefont {J.-c.}\ \bibnamefont
  {Hwang}},\ }\href {\doibase 10.1103/PhysRevD.69.104011} {\bibfield  {journal}
  {\bibinfo  {journal} {Phys. Rev.}\ }\textbf {\bibinfo {volume} {D69}},\
  \bibinfo {pages} {104011} (\bibinfo {year} {2004})}\BibitemShut {NoStop}%
\bibitem [{\citenamefont {Carbone}\ and\ \citenamefont
  {Matarrese}(2005)}]{Carbone:2004iv}%
  \BibitemOpen
  \bibfield  {author} {\bibinfo {author} {\bibfnamefont {C.}~\bibnamefont
  {Carbone}}\ and\ \bibinfo {author} {\bibfnamefont {S.}~\bibnamefont
  {Matarrese}},\ }\href {\doibase 10.1103/PhysRevD.71.043508} {\bibfield
  {journal} {\bibinfo  {journal} {Phys. Rev.}\ }\textbf {\bibinfo {volume}
  {D71}},\ \bibinfo {pages} {043508} (\bibinfo {year} {2005})},\ \Eprint
  {http://arxiv.org/abs/astro-ph/0407611} {arXiv:astro-ph/0407611 [astro-ph]}
  \BibitemShut {NoStop}%
\bibitem [{\citenamefont {Nakamura}(2007)}]{Nakamura:2004rm}%
  \BibitemOpen
  \bibfield  {author} {\bibinfo {author} {\bibfnamefont {K.}~\bibnamefont
  {Nakamura}},\ }\href {\doibase 10.1143/PTP.117.17} {\bibfield  {journal}
  {\bibinfo  {journal} {Prog. Theor. Phys.}\ }\textbf {\bibinfo {volume}
  {117}},\ \bibinfo {pages} {17} (\bibinfo {year} {2007})},\ \Eprint
  {http://arxiv.org/abs/gr-qc/0605108} {arXiv:gr-qc/0605108 [gr-qc]}
  \BibitemShut {NoStop}%
\bibitem [{\citenamefont {Yuan}\ \emph
  {et~al.}(2019{\natexlab{a}})\citenamefont {Yuan}, \citenamefont {Chen},\ and\
  \citenamefont {Huang}}]{Yuan:2019udt}%
  \BibitemOpen
  \bibfield  {author} {\bibinfo {author} {\bibfnamefont {C.}~\bibnamefont
  {Yuan}}, \bibinfo {author} {\bibfnamefont {Z.-C.}\ \bibnamefont {Chen}}, \
  and\ \bibinfo {author} {\bibfnamefont {Q.-G.}\ \bibnamefont {Huang}},\ }\href
  {\doibase 10.1103/PhysRevD.100.081301} {\bibfield  {journal} {\bibinfo
  {journal} {Phys. Rev.}\ }\textbf {\bibinfo {volume} {D100}},\ \bibinfo
  {pages} {081301} (\bibinfo {year} {2019}{\natexlab{a}})},\ \Eprint
  {http://arxiv.org/abs/1906.11549} {arXiv:1906.11549 [astro-ph.CO]}
  \BibitemShut {NoStop}%
\bibitem [{\citenamefont {Ananda}\ \emph {et~al.}(2007)\citenamefont {Ananda},
  \citenamefont {Clarkson},\ and\ \citenamefont {Wands}}]{Ananda:2006af}%
  \BibitemOpen
  \bibfield  {author} {\bibinfo {author} {\bibfnamefont {K.~N.}\ \bibnamefont
  {Ananda}}, \bibinfo {author} {\bibfnamefont {C.}~\bibnamefont {Clarkson}}, \
  and\ \bibinfo {author} {\bibfnamefont {D.}~\bibnamefont {Wands}},\ }\href
  {\doibase 10.1103/PhysRevD.75.123518} {\bibfield  {journal} {\bibinfo
  {journal} {Phys. Rev.}\ }\textbf {\bibinfo {volume} {D75}},\ \bibinfo {pages}
  {123518} (\bibinfo {year} {2007})},\ \Eprint
  {http://arxiv.org/abs/gr-qc/0612013} {arXiv:gr-qc/0612013 [gr-qc]}
  \BibitemShut {NoStop}%
\bibitem [{\citenamefont {Baumann}\ \emph {et~al.}(2007)\citenamefont
  {Baumann}, \citenamefont {Steinhardt}, \citenamefont {Takahashi},\ and\
  \citenamefont {Ichiki}}]{Baumann:2007zm}%
  \BibitemOpen
  \bibfield  {author} {\bibinfo {author} {\bibfnamefont {D.}~\bibnamefont
  {Baumann}}, \bibinfo {author} {\bibfnamefont {P.~J.}\ \bibnamefont
  {Steinhardt}}, \bibinfo {author} {\bibfnamefont {K.}~\bibnamefont
  {Takahashi}}, \ and\ \bibinfo {author} {\bibfnamefont {K.}~\bibnamefont
  {Ichiki}},\ }\href {\doibase 10.1103/PhysRevD.76.084019} {\bibfield
  {journal} {\bibinfo  {journal} {Phys. Rev.}\ }\textbf {\bibinfo {volume}
  {D76}},\ \bibinfo {pages} {084019} (\bibinfo {year} {2007})},\ \Eprint
  {http://arxiv.org/abs/hep-th/0703290} {arXiv:hep-th/0703290 [hep-th]}
  \BibitemShut {NoStop}%
\bibitem [{\citenamefont {Saito}\ and\ \citenamefont
  {Yokoyama}(2009)}]{Saito:2008jc}%
  \BibitemOpen
  \bibfield  {author} {\bibinfo {author} {\bibfnamefont {R.}~\bibnamefont
  {Saito}}\ and\ \bibinfo {author} {\bibfnamefont {J.}~\bibnamefont
  {Yokoyama}},\ }\href {\doibase 10.1103/PhysRevLett.102.161101,
  10.1103/PhysRevLett.107.069901} {\bibfield  {journal} {\bibinfo  {journal}
  {Phys. Rev. Lett.}\ }\textbf {\bibinfo {volume} {102}},\ \bibinfo {pages}
  {161101} (\bibinfo {year} {2009})},\ \bibinfo {note} {[Erratum: Phys. Rev.
  Lett.107,069901(2011)]},\ \Eprint {http://arxiv.org/abs/0812.4339}
  {arXiv:0812.4339 [astro-ph]} \BibitemShut {NoStop}%
\bibitem [{\citenamefont {Assadullahi}\ and\ \citenamefont
  {Wands}(2010)}]{Assadullahi:2009jc}%
  \BibitemOpen
  \bibfield  {author} {\bibinfo {author} {\bibfnamefont {H.}~\bibnamefont
  {Assadullahi}}\ and\ \bibinfo {author} {\bibfnamefont {D.}~\bibnamefont
  {Wands}},\ }\href {\doibase 10.1103/PhysRevD.81.023527} {\bibfield  {journal}
  {\bibinfo  {journal} {Phys. Rev.}\ }\textbf {\bibinfo {volume} {D81}},\
  \bibinfo {pages} {023527} (\bibinfo {year} {2010})},\ \Eprint
  {http://arxiv.org/abs/0907.4073} {arXiv:0907.4073 [astro-ph.CO]} \BibitemShut
  {NoStop}%
\bibitem [{\citenamefont {Bugaev}\ and\ \citenamefont
  {Klimai}(2010)}]{Bugaev:2009zh}%
  \BibitemOpen
  \bibfield  {author} {\bibinfo {author} {\bibfnamefont {E.}~\bibnamefont
  {Bugaev}}\ and\ \bibinfo {author} {\bibfnamefont {P.}~\bibnamefont
  {Klimai}},\ }\href {\doibase 10.1103/PhysRevD.81.023517} {\bibfield
  {journal} {\bibinfo  {journal} {Phys. Rev.}\ }\textbf {\bibinfo {volume}
  {D81}},\ \bibinfo {pages} {023517} (\bibinfo {year} {2010})},\ \Eprint
  {http://arxiv.org/abs/0908.0664} {arXiv:0908.0664 [astro-ph.CO]} \BibitemShut
  {NoStop}%
\bibitem [{\citenamefont {Saito}\ and\ \citenamefont
  {Yokoyama}(2010)}]{Saito:2009jt}%
  \BibitemOpen
  \bibfield  {author} {\bibinfo {author} {\bibfnamefont {R.}~\bibnamefont
  {Saito}}\ and\ \bibinfo {author} {\bibfnamefont {J.}~\bibnamefont
  {Yokoyama}},\ }\href {\doibase 10.1143/PTP.126.351, 10.1143/PTP.123.867}
  {\bibfield  {journal} {\bibinfo  {journal} {Prog. Theor. Phys.}\ }\textbf
  {\bibinfo {volume} {123}},\ \bibinfo {pages} {867} (\bibinfo {year}
  {2010})},\ \bibinfo {note} {[Erratum: Prog. Theor. Phys.126,351(2011)]},\
  \Eprint {http://arxiv.org/abs/0912.5317} {arXiv:0912.5317 [astro-ph.CO]}
  \BibitemShut {NoStop}%
\bibitem [{\citenamefont {Bugaev}\ and\ \citenamefont
  {Klimai}(2011)}]{Bugaev:2010bb}%
  \BibitemOpen
  \bibfield  {author} {\bibinfo {author} {\bibfnamefont {E.}~\bibnamefont
  {Bugaev}}\ and\ \bibinfo {author} {\bibfnamefont {P.}~\bibnamefont
  {Klimai}},\ }\href {\doibase 10.1103/PhysRevD.83.083521} {\bibfield
  {journal} {\bibinfo  {journal} {Phys. Rev.}\ }\textbf {\bibinfo {volume}
  {D83}},\ \bibinfo {pages} {083521} (\bibinfo {year} {2011})},\ \Eprint
  {http://arxiv.org/abs/1012.4697} {arXiv:1012.4697 [astro-ph.CO]} \BibitemShut
  {NoStop}%
\bibitem [{\citenamefont {Nakama}\ and\ \citenamefont
  {Suyama}(2016)}]{Nakama:2016enz}%
  \BibitemOpen
  \bibfield  {author} {\bibinfo {author} {\bibfnamefont {T.}~\bibnamefont
  {Nakama}}\ and\ \bibinfo {author} {\bibfnamefont {T.}~\bibnamefont
  {Suyama}},\ }\href {\doibase 10.1103/PhysRevD.94.043507} {\bibfield
  {journal} {\bibinfo  {journal} {Phys. Rev.}\ }\textbf {\bibinfo {volume}
  {D94}},\ \bibinfo {pages} {043507} (\bibinfo {year} {2016})},\ \Eprint
  {http://arxiv.org/abs/1605.04482} {arXiv:1605.04482 [gr-qc]} \BibitemShut
  {NoStop}%
\bibitem [{\citenamefont {Nakama}\ \emph {et~al.}(2017)\citenamefont {Nakama},
  \citenamefont {Silk},\ and\ \citenamefont {Kamionkowski}}]{Nakama:2016gzw}%
  \BibitemOpen
  \bibfield  {author} {\bibinfo {author} {\bibfnamefont {T.}~\bibnamefont
  {Nakama}}, \bibinfo {author} {\bibfnamefont {J.}~\bibnamefont {Silk}}, \ and\
  \bibinfo {author} {\bibfnamefont {M.}~\bibnamefont {Kamionkowski}},\ }\href
  {\doibase 10.1103/PhysRevD.95.043511} {\bibfield  {journal} {\bibinfo
  {journal} {Phys. Rev.}\ }\textbf {\bibinfo {volume} {D95}},\ \bibinfo {pages}
  {043511} (\bibinfo {year} {2017})},\ \Eprint
  {http://arxiv.org/abs/1612.06264} {arXiv:1612.06264 [astro-ph.CO]}
  \BibitemShut {NoStop}%
\bibitem [{\citenamefont {Garcia-Bellido}\ \emph {et~al.}(2017)\citenamefont
  {Garcia-Bellido}, \citenamefont {Peloso},\ and\ \citenamefont
  {Unal}}]{Garcia-Bellido:2017aan}%
  \BibitemOpen
  \bibfield  {author} {\bibinfo {author} {\bibfnamefont {J.}~\bibnamefont
  {Garcia-Bellido}}, \bibinfo {author} {\bibfnamefont {M.}~\bibnamefont
  {Peloso}}, \ and\ \bibinfo {author} {\bibfnamefont {C.}~\bibnamefont
  {Unal}},\ }\href {\doibase 10.1088/1475-7516/2017/09/013} {\bibfield
  {journal} {\bibinfo  {journal} {JCAP}\ }\textbf {\bibinfo {volume} {1709}},\
  \bibinfo {pages} {013} (\bibinfo {year} {2017})},\ \Eprint
  {http://arxiv.org/abs/1707.02441} {arXiv:1707.02441 [astro-ph.CO]}
  \BibitemShut {NoStop}%
\bibitem [{\citenamefont {Sasaki}\ \emph {et~al.}(2018)\citenamefont {Sasaki},
  \citenamefont {Suyama}, \citenamefont {Tanaka},\ and\ \citenamefont
  {Yokoyama}}]{Sasaki:2018dmp}%
  \BibitemOpen
  \bibfield  {author} {\bibinfo {author} {\bibfnamefont {M.}~\bibnamefont
  {Sasaki}}, \bibinfo {author} {\bibfnamefont {T.}~\bibnamefont {Suyama}},
  \bibinfo {author} {\bibfnamefont {T.}~\bibnamefont {Tanaka}}, \ and\ \bibinfo
  {author} {\bibfnamefont {S.}~\bibnamefont {Yokoyama}},\ }\href {\doibase
  10.1088/1361-6382/aaa7b4} {\bibfield  {journal} {\bibinfo  {journal} {Class.
  Quant. Grav.}\ }\textbf {\bibinfo {volume} {35}},\ \bibinfo {pages} {063001}
  (\bibinfo {year} {2018})},\ \Eprint {http://arxiv.org/abs/1801.05235}
  {arXiv:1801.05235 [astro-ph.CO]} \BibitemShut {NoStop}%
\bibitem [{\citenamefont {Espinosa}\ \emph {et~al.}(2018)\citenamefont
  {Espinosa}, \citenamefont {Racco},\ and\ \citenamefont
  {Riotto}}]{Espinosa:2018eve}%
  \BibitemOpen
  \bibfield  {author} {\bibinfo {author} {\bibfnamefont {J.~R.}\ \bibnamefont
  {Espinosa}}, \bibinfo {author} {\bibfnamefont {D.}~\bibnamefont {Racco}}, \
  and\ \bibinfo {author} {\bibfnamefont {A.}~\bibnamefont {Riotto}},\ }\href
  {\doibase 10.1088/1475-7516/2018/09/012} {\bibfield  {journal} {\bibinfo
  {journal} {JCAP}\ }\textbf {\bibinfo {volume} {1809}},\ \bibinfo {pages}
  {012} (\bibinfo {year} {2018})},\ \Eprint {http://arxiv.org/abs/1804.07732}
  {arXiv:1804.07732 [hep-ph]} \BibitemShut {NoStop}%
\bibitem [{\citenamefont {Kohri}\ and\ \citenamefont
  {Terada}(2018)}]{Kohri:2018awv}%
  \BibitemOpen
  \bibfield  {author} {\bibinfo {author} {\bibfnamefont {K.}~\bibnamefont
  {Kohri}}\ and\ \bibinfo {author} {\bibfnamefont {T.}~\bibnamefont {Terada}},\
  }\href {\doibase 10.1103/PhysRevD.97.123532} {\bibfield  {journal} {\bibinfo
  {journal} {Phys. Rev.}\ }\textbf {\bibinfo {volume} {D97}},\ \bibinfo {pages}
  {123532} (\bibinfo {year} {2018})},\ \Eprint
  {http://arxiv.org/abs/1804.08577} {arXiv:1804.08577 [gr-qc]} \BibitemShut
  {NoStop}%
\bibitem [{\citenamefont {Cai}\ \emph {et~al.}(2019{\natexlab{a}})\citenamefont
  {Cai}, \citenamefont {Pi},\ and\ \citenamefont {Sasaki}}]{Cai:2018dig}%
  \BibitemOpen
  \bibfield  {author} {\bibinfo {author} {\bibfnamefont {R.-g.}\ \bibnamefont
  {Cai}}, \bibinfo {author} {\bibfnamefont {S.}~\bibnamefont {Pi}}, \ and\
  \bibinfo {author} {\bibfnamefont {M.}~\bibnamefont {Sasaki}},\ }\href
  {\doibase 10.1103/PhysRevLett.122.201101} {\bibfield  {journal} {\bibinfo
  {journal} {Phys. Rev. Lett.}\ }\textbf {\bibinfo {volume} {122}},\ \bibinfo
  {pages} {201101} (\bibinfo {year} {2019}{\natexlab{a}})},\ \Eprint
  {http://arxiv.org/abs/1810.11000} {arXiv:1810.11000 [astro-ph.CO]}
  \BibitemShut {NoStop}%
\bibitem [{\citenamefont {Bartolo}\ \emph
  {et~al.}(2019{\natexlab{a}})\citenamefont {Bartolo}, \citenamefont {De~Luca},
  \citenamefont {Franciolini}, \citenamefont {Lewis}, \citenamefont {Peloso},\
  and\ \citenamefont {Riotto}}]{Bartolo:2018evs}%
  \BibitemOpen
  \bibfield  {author} {\bibinfo {author} {\bibfnamefont {N.}~\bibnamefont
  {Bartolo}}, \bibinfo {author} {\bibfnamefont {V.}~\bibnamefont {De~Luca}},
  \bibinfo {author} {\bibfnamefont {G.}~\bibnamefont {Franciolini}}, \bibinfo
  {author} {\bibfnamefont {A.}~\bibnamefont {Lewis}}, \bibinfo {author}
  {\bibfnamefont {M.}~\bibnamefont {Peloso}}, \ and\ \bibinfo {author}
  {\bibfnamefont {A.}~\bibnamefont {Riotto}},\ }\href {\doibase
  10.1103/PhysRevLett.122.211301} {\bibfield  {journal} {\bibinfo  {journal}
  {Phys. Rev. Lett.}\ }\textbf {\bibinfo {volume} {122}},\ \bibinfo {pages}
  {211301} (\bibinfo {year} {2019}{\natexlab{a}})},\ \Eprint
  {http://arxiv.org/abs/1810.12218} {arXiv:1810.12218 [astro-ph.CO]}
  \BibitemShut {NoStop}%
\bibitem [{\citenamefont {Bartolo}\ \emph
  {et~al.}(2019{\natexlab{b}})\citenamefont {Bartolo}, \citenamefont {De~Luca},
  \citenamefont {Franciolini}, \citenamefont {Peloso}, \citenamefont {Racco},\
  and\ \citenamefont {Riotto}}]{Bartolo:2018rku}%
  \BibitemOpen
  \bibfield  {author} {\bibinfo {author} {\bibfnamefont {N.}~\bibnamefont
  {Bartolo}}, \bibinfo {author} {\bibfnamefont {V.}~\bibnamefont {De~Luca}},
  \bibinfo {author} {\bibfnamefont {G.}~\bibnamefont {Franciolini}}, \bibinfo
  {author} {\bibfnamefont {M.}~\bibnamefont {Peloso}}, \bibinfo {author}
  {\bibfnamefont {D.}~\bibnamefont {Racco}}, \ and\ \bibinfo {author}
  {\bibfnamefont {A.}~\bibnamefont {Riotto}},\ }\href {\doibase
  10.1103/PhysRevD.99.103521} {\bibfield  {journal} {\bibinfo  {journal} {Phys.
  Rev.}\ }\textbf {\bibinfo {volume} {D99}},\ \bibinfo {pages} {103521}
  (\bibinfo {year} {2019}{\natexlab{b}})},\ \Eprint
  {http://arxiv.org/abs/1810.12224} {arXiv:1810.12224 [astro-ph.CO]}
  \BibitemShut {NoStop}%
\bibitem [{\citenamefont {Unal}(2019)}]{Unal:2018yaa}%
  \BibitemOpen
  \bibfield  {author} {\bibinfo {author} {\bibfnamefont {C.}~\bibnamefont
  {Unal}},\ }\href {\doibase 10.1103/PhysRevD.99.041301} {\bibfield  {journal}
  {\bibinfo  {journal} {Phys. Rev.}\ }\textbf {\bibinfo {volume} {D99}},\
  \bibinfo {pages} {041301} (\bibinfo {year} {2019})},\ \Eprint
  {http://arxiv.org/abs/1811.09151} {arXiv:1811.09151 [astro-ph.CO]}
  \BibitemShut {NoStop}%
\bibitem [{\citenamefont {Inomata}\ and\ \citenamefont
  {Nakama}(2019)}]{Inomata:2018epa}%
  \BibitemOpen
  \bibfield  {author} {\bibinfo {author} {\bibfnamefont {K.}~\bibnamefont
  {Inomata}}\ and\ \bibinfo {author} {\bibfnamefont {T.}~\bibnamefont
  {Nakama}},\ }\href {\doibase 10.1103/PhysRevD.99.043511} {\bibfield
  {journal} {\bibinfo  {journal} {Phys. Rev.}\ }\textbf {\bibinfo {volume}
  {D99}},\ \bibinfo {pages} {043511} (\bibinfo {year} {2019})},\ \Eprint
  {http://arxiv.org/abs/1812.00674} {arXiv:1812.00674 [astro-ph.CO]}
  \BibitemShut {NoStop}%
\bibitem [{\citenamefont {Clesse}\ \emph {et~al.}(2018)\citenamefont {Clesse},
  \citenamefont {García-Bellido},\ and\ \citenamefont
  {Orani}}]{Clesse:2018ogk}%
  \BibitemOpen
  \bibfield  {author} {\bibinfo {author} {\bibfnamefont {S.}~\bibnamefont
  {Clesse}}, \bibinfo {author} {\bibfnamefont {J.}~\bibnamefont
  {García-Bellido}}, \ and\ \bibinfo {author} {\bibfnamefont {S.}~\bibnamefont
  {Orani}},\ }\href@noop {} {\  (\bibinfo {year} {2018})},\ \Eprint
  {http://arxiv.org/abs/1812.11011} {arXiv:1812.11011 [astro-ph.CO]}
  \BibitemShut {NoStop}%
\bibitem [{\citenamefont {Cai}\ \emph {et~al.}(2019{\natexlab{b}})\citenamefont
  {Cai}, \citenamefont {Pi}, \citenamefont {Wang},\ and\ \citenamefont
  {Yang}}]{Cai:2019amo}%
  \BibitemOpen
  \bibfield  {author} {\bibinfo {author} {\bibfnamefont {R.-G.}\ \bibnamefont
  {Cai}}, \bibinfo {author} {\bibfnamefont {S.}~\bibnamefont {Pi}}, \bibinfo
  {author} {\bibfnamefont {S.-J.}\ \bibnamefont {Wang}}, \ and\ \bibinfo
  {author} {\bibfnamefont {X.-Y.}\ \bibnamefont {Yang}},\ }\href {\doibase
  10.1088/1475-7516/2019/05/013} {\  (\bibinfo {year} {2019}{\natexlab{b}}),\
  10.1088/1475-7516/2019/05/013},\ \bibinfo {note}
  {[JCAP1905,no.05,013(2019)]},\ \Eprint {http://arxiv.org/abs/1901.10152}
  {arXiv:1901.10152 [astro-ph.CO]} \BibitemShut {NoStop}%
\bibitem [{\citenamefont {Inomata}\ \emph
  {et~al.}(2019{\natexlab{a}})\citenamefont {Inomata}, \citenamefont {Kohri},
  \citenamefont {Nakama},\ and\ \citenamefont {Terada}}]{Inomata:2019zqy}%
  \BibitemOpen
  \bibfield  {author} {\bibinfo {author} {\bibfnamefont {K.}~\bibnamefont
  {Inomata}}, \bibinfo {author} {\bibfnamefont {K.}~\bibnamefont {Kohri}},
  \bibinfo {author} {\bibfnamefont {T.}~\bibnamefont {Nakama}}, \ and\ \bibinfo
  {author} {\bibfnamefont {T.}~\bibnamefont {Terada}},\ }\href@noop {} {\
  (\bibinfo {year} {2019}{\natexlab{a}})},\ \Eprint
  {http://arxiv.org/abs/1904.12878} {arXiv:1904.12878 [astro-ph.CO]}
  \BibitemShut {NoStop}%
\bibitem [{\citenamefont {Inomata}\ \emph
  {et~al.}(2019{\natexlab{b}})\citenamefont {Inomata}, \citenamefont {Kohri},
  \citenamefont {Nakama},\ and\ \citenamefont {Terada}}]{Inomata:2019ivs}%
  \BibitemOpen
  \bibfield  {author} {\bibinfo {author} {\bibfnamefont {K.}~\bibnamefont
  {Inomata}}, \bibinfo {author} {\bibfnamefont {K.}~\bibnamefont {Kohri}},
  \bibinfo {author} {\bibfnamefont {T.}~\bibnamefont {Nakama}}, \ and\ \bibinfo
  {author} {\bibfnamefont {T.}~\bibnamefont {Terada}},\ }\href {\doibase
  10.1103/PhysRevD.100.043532} {\bibfield  {journal} {\bibinfo  {journal}
  {Phys. Rev.}\ }\textbf {\bibinfo {volume} {D100}},\ \bibinfo {pages} {043532}
  (\bibinfo {year} {2019}{\natexlab{b}})},\ \Eprint
  {http://arxiv.org/abs/1904.12879} {arXiv:1904.12879 [astro-ph.CO]}
  \BibitemShut {NoStop}%
\bibitem [{\citenamefont {Cai}\ \emph {et~al.}(2019{\natexlab{c}})\citenamefont
  {Cai}, \citenamefont {Pi}, \citenamefont {Wang},\ and\ \citenamefont
  {Yang}}]{Cai:2019elf}%
  \BibitemOpen
  \bibfield  {author} {\bibinfo {author} {\bibfnamefont {R.-G.}\ \bibnamefont
  {Cai}}, \bibinfo {author} {\bibfnamefont {S.}~\bibnamefont {Pi}}, \bibinfo
  {author} {\bibfnamefont {S.-J.}\ \bibnamefont {Wang}}, \ and\ \bibinfo
  {author} {\bibfnamefont {X.-Y.}\ \bibnamefont {Yang}},\ }\href@noop {} {\
  (\bibinfo {year} {2019}{\natexlab{c}})},\ \Eprint
  {http://arxiv.org/abs/1907.06372} {arXiv:1907.06372 [astro-ph.CO]}
  \BibitemShut {NoStop}%
\bibitem [{\citenamefont {Cai}\ \emph {et~al.}(2019{\natexlab{d}})\citenamefont
  {Cai}, \citenamefont {Pi},\ and\ \citenamefont {Sasaki}}]{Cai:2019cdl}%
  \BibitemOpen
  \bibfield  {author} {\bibinfo {author} {\bibfnamefont {R.-G.}\ \bibnamefont
  {Cai}}, \bibinfo {author} {\bibfnamefont {S.}~\bibnamefont {Pi}}, \ and\
  \bibinfo {author} {\bibfnamefont {M.}~\bibnamefont {Sasaki}},\ }\href@noop {}
  {\  (\bibinfo {year} {2019}{\natexlab{d}})},\ \Eprint
  {http://arxiv.org/abs/1909.13728} {arXiv:1909.13728 [astro-ph.CO]}
  \BibitemShut {NoStop}%
\bibitem [{\citenamefont {Chen}\ \emph
  {et~al.}(2019{\natexlab{b}})\citenamefont {Chen}, \citenamefont {Yuan},\ and\
  \citenamefont {Huang}}]{Chen:2019xse}%
  \BibitemOpen
  \bibfield  {author} {\bibinfo {author} {\bibfnamefont {Z.-C.}\ \bibnamefont
  {Chen}}, \bibinfo {author} {\bibfnamefont {C.}~\bibnamefont {Yuan}}, \ and\
  \bibinfo {author} {\bibfnamefont {Q.-G.}\ \bibnamefont {Huang}},\ }\href@noop
  {} {\  (\bibinfo {year} {2019}{\natexlab{b}})},\ \Eprint
  {http://arxiv.org/abs/1910.12239} {arXiv:1910.12239 [astro-ph.CO]}
  \BibitemShut {NoStop}%
\bibitem [{\citenamefont {Yuan}\ \emph
  {et~al.}(2019{\natexlab{b}})\citenamefont {Yuan}, \citenamefont {Chen},\ and\
  \citenamefont {Huang}}]{Yuan:2019fwv}%
  \BibitemOpen
  \bibfield  {author} {\bibinfo {author} {\bibfnamefont {C.}~\bibnamefont
  {Yuan}}, \bibinfo {author} {\bibfnamefont {Z.-C.}\ \bibnamefont {Chen}}, \
  and\ \bibinfo {author} {\bibfnamefont {Q.-G.}\ \bibnamefont {Huang}},\
  }\href@noop {} {\  (\bibinfo {year} {2019}{\natexlab{b}})},\ \Eprint
  {http://arxiv.org/abs/1912.00885} {arXiv:1912.00885 [astro-ph.CO]}
  \BibitemShut {NoStop}%
\bibitem [{\citenamefont {Maggiore}(2000)}]{Maggiore:1999vm}%
  \BibitemOpen
  \bibfield  {author} {\bibinfo {author} {\bibfnamefont {M.}~\bibnamefont
  {Maggiore}},\ }\href {\doibase 10.1016/S0370-1573(99)00102-7} {\bibfield
  {journal} {\bibinfo  {journal} {Phys. Rept.}\ }\textbf {\bibinfo {volume}
  {331}},\ \bibinfo {pages} {283} (\bibinfo {year} {2000})},\ \Eprint
  {http://arxiv.org/abs/gr-qc/9909001} {arXiv:gr-qc/9909001 [gr-qc]}
  \BibitemShut {NoStop}%
\bibitem [{\citenamefont {Thrane}\ and\ \citenamefont
  {Romano}(2013)}]{Thrane:2013oya}%
  \BibitemOpen
  \bibfield  {author} {\bibinfo {author} {\bibfnamefont {E.}~\bibnamefont
  {Thrane}}\ and\ \bibinfo {author} {\bibfnamefont {J.~D.}\ \bibnamefont
  {Romano}},\ }\href {\doibase 10.1103/PhysRevD.88.124032} {\bibfield
  {journal} {\bibinfo  {journal} {Phys. Rev.}\ }\textbf {\bibinfo {volume}
  {D88}},\ \bibinfo {pages} {124032} (\bibinfo {year} {2013})},\ \Eprint
  {http://arxiv.org/abs/1310.5300} {arXiv:1310.5300 [astro-ph.IM]} \BibitemShut
  {NoStop}%
\bibitem [{\citenamefont {Lasky}\ \emph {et~al.}(2016)\citenamefont {Lasky}
  \emph {et~al.}}]{Lasky:2015lej}%
  \BibitemOpen
  \bibfield  {author} {\bibinfo {author} {\bibfnamefont {P.~D.}\ \bibnamefont
  {Lasky}} \emph {et~al.},\ }\href {\doibase 10.1103/PhysRevX.6.011035}
  {\bibfield  {journal} {\bibinfo  {journal} {Phys. Rev.}\ }\textbf {\bibinfo
  {volume} {X6}},\ \bibinfo {pages} {011035} (\bibinfo {year} {2016})},\
  \Eprint {http://arxiv.org/abs/1511.05994} {arXiv:1511.05994 [astro-ph.CO]}
  \BibitemShut {NoStop}%
\bibitem [{\citenamefont {Li}\ \emph {et~al.}(2019)\citenamefont {Li},
  \citenamefont {Chen},\ and\ \citenamefont {Huang}}]{Li:2019vlb}%
  \BibitemOpen
  \bibfield  {author} {\bibinfo {author} {\bibfnamefont {J.}~\bibnamefont
  {Li}}, \bibinfo {author} {\bibfnamefont {Z.-C.}\ \bibnamefont {Chen}}, \ and\
  \bibinfo {author} {\bibfnamefont {Q.-G.}\ \bibnamefont {Huang}},\ }\href
  {\doibase 10.1007/s11433-019-9605-5} {\bibfield  {journal} {\bibinfo
  {journal} {Sci. China Phys. Mech. Astron.}\ }\textbf {\bibinfo {volume}
  {62}},\ \bibinfo {pages} {110421} (\bibinfo {year} {2019})},\ \Eprint
  {http://arxiv.org/abs/1907.09794} {arXiv:1907.09794 [astro-ph.CO]}
  \BibitemShut {NoStop}%
\bibitem [{\citenamefont {Li}\ and\ \citenamefont {Huang}(2019)}]{Li:2019efi}%
  \BibitemOpen
  \bibfield  {author} {\bibinfo {author} {\bibfnamefont {J.}~\bibnamefont
  {Li}}\ and\ \bibinfo {author} {\bibfnamefont {Q.-G.}\ \bibnamefont {Huang}},\
  }\href {\doibase 10.1007/s11433-019-9446-1} {\bibfield  {journal} {\bibinfo
  {journal} {Sci. China Phys. Mech. Astron.}\ }\textbf {\bibinfo {volume}
  {62}},\ \bibinfo {pages} {120412} (\bibinfo {year} {2019})},\ \Eprint
  {http://arxiv.org/abs/1906.01336} {arXiv:1906.01336 [astro-ph.CO]}
  \BibitemShut {NoStop}%
\bibitem [{\citenamefont {Li}\ and\ \citenamefont {Huang}(2018)}]{Li:2018iwg}%
  \BibitemOpen
  \bibfield  {author} {\bibinfo {author} {\bibfnamefont {J.}~\bibnamefont
  {Li}}\ and\ \bibinfo {author} {\bibfnamefont {Q.-G.}\ \bibnamefont {Huang}},\
  }\href {\doibase 10.1140/epjc/s10052-018-6471-z} {\bibfield  {journal}
  {\bibinfo  {journal} {Eur. Phys. J.}\ }\textbf {\bibinfo {volume} {C78}},\
  \bibinfo {pages} {980} (\bibinfo {year} {2018})},\ \Eprint
  {http://arxiv.org/abs/1806.01440} {arXiv:1806.01440 [astro-ph.CO]}
  \BibitemShut {NoStop}%
\bibitem [{\citenamefont {Caprini}\ \emph {et~al.}(2009)\citenamefont
  {Caprini}, \citenamefont {Durrer}, \citenamefont {Konstandin},\ and\
  \citenamefont {Servant}}]{Caprini:2009fx}%
  \BibitemOpen
  \bibfield  {author} {\bibinfo {author} {\bibfnamefont {C.}~\bibnamefont
  {Caprini}}, \bibinfo {author} {\bibfnamefont {R.}~\bibnamefont {Durrer}},
  \bibinfo {author} {\bibfnamefont {T.}~\bibnamefont {Konstandin}}, \ and\
  \bibinfo {author} {\bibfnamefont {G.}~\bibnamefont {Servant}},\ }\href
  {\doibase 10.1103/PhysRevD.79.083519} {\bibfield  {journal} {\bibinfo
  {journal} {Phys. Rev.}\ }\textbf {\bibinfo {volume} {D79}},\ \bibinfo {pages}
  {083519} (\bibinfo {year} {2009})},\ \Eprint {http://arxiv.org/abs/0901.1661}
  {arXiv:0901.1661 [astro-ph.CO]} \BibitemShut {NoStop}%
\bibitem [{\citenamefont {Liddle}\ \emph {et~al.}(2000)\citenamefont {Liddle},
  \citenamefont {Lyth}, \citenamefont {Malik},\ and\ \citenamefont
  {Wands}}]{Liddle:1999hq}%
  \BibitemOpen
  \bibfield  {author} {\bibinfo {author} {\bibfnamefont {A.~R.}\ \bibnamefont
  {Liddle}}, \bibinfo {author} {\bibfnamefont {D.~H.}\ \bibnamefont {Lyth}},
  \bibinfo {author} {\bibfnamefont {K.~A.}\ \bibnamefont {Malik}}, \ and\
  \bibinfo {author} {\bibfnamefont {D.}~\bibnamefont {Wands}},\ }\href
  {\doibase 10.1103/PhysRevD.61.103509} {\bibfield  {journal} {\bibinfo
  {journal} {Phys. Rev.}\ }\textbf {\bibinfo {volume} {D61}},\ \bibinfo {pages}
  {103509} (\bibinfo {year} {2000})},\ \Eprint
  {http://arxiv.org/abs/hep-ph/9912473} {arXiv:hep-ph/9912473 [hep-ph]}
  \BibitemShut {NoStop}%
\bibitem [{\citenamefont {Audley}\ \emph {et~al.}(2017)\citenamefont {Audley}
  \emph {et~al.}}]{Audley:2017drz}%
  \BibitemOpen
  \bibfield  {author} {\bibinfo {author} {\bibfnamefont {H.}~\bibnamefont
  {Audley}} \emph {et~al.} (\bibinfo {collaboration} {LISA}),\ }\href@noop {}
  {\  (\bibinfo {year} {2017})},\ \Eprint {http://arxiv.org/abs/1702.00786}
  {arXiv:1702.00786 [astro-ph.IM]} \BibitemShut {NoStop}%
\bibitem [{\citenamefont {Kuroyanagi}\ \emph {et~al.}(2018)\citenamefont
  {Kuroyanagi}, \citenamefont {Chiba},\ and\ \citenamefont
  {Takahashi}}]{Kuroyanagi:2018csn}%
  \BibitemOpen
  \bibfield  {author} {\bibinfo {author} {\bibfnamefont {S.}~\bibnamefont
  {Kuroyanagi}}, \bibinfo {author} {\bibfnamefont {T.}~\bibnamefont {Chiba}}, \
  and\ \bibinfo {author} {\bibfnamefont {T.}~\bibnamefont {Takahashi}},\ }\href
  {\doibase 10.1088/1475-7516/2018/11/038} {\bibfield  {journal} {\bibinfo
  {journal} {JCAP}\ }\textbf {\bibinfo {volume} {1811}},\ \bibinfo {pages}
  {038} (\bibinfo {year} {2018})},\ \Eprint {http://arxiv.org/abs/1807.00786}
  {arXiv:1807.00786 [astro-ph.CO]} \BibitemShut {NoStop}%
\bibitem [{\citenamefont {Robson}\ \emph {et~al.}(2019)\citenamefont {Robson},
  \citenamefont {Cornish},\ and\ \citenamefont {Liu}}]{Cornish:2018dyw}%
  \BibitemOpen
  \bibfield  {author} {\bibinfo {author} {\bibfnamefont {T.}~\bibnamefont
  {Robson}}, \bibinfo {author} {\bibfnamefont {N.~J.}\ \bibnamefont {Cornish}},
  \ and\ \bibinfo {author} {\bibfnamefont {C.}~\bibnamefont {Liu}},\ }\href
  {\doibase 10.1088/1361-6382/ab1101} {\bibfield  {journal} {\bibinfo
  {journal} {Class. Quant. Grav.}\ }\textbf {\bibinfo {volume} {36}},\ \bibinfo
  {pages} {105011} (\bibinfo {year} {2019})},\ \Eprint
  {http://arxiv.org/abs/1803.01944} {arXiv:1803.01944 [astro-ph.HE]}
  \BibitemShut {NoStop}%
\end{thebibliography}%
	
\end{document}